%% file: main_clean.tex
\documentclass[english,notitlepage,superscriptaddress,nofootinbib,twocolumn,aps,pra,10pt]{revtex4-2}

\usepackage[T1]{fontenc}
\usepackage[utf8]{inputenc}
\usepackage[english]{babel}
\usepackage{makecell}
\usepackage{amsthm,amsmath,amssymb}
\usepackage{graphicx}
\usepackage{braket}
\usepackage{multirow}
\usepackage{lipsum}
\usepackage{dsfont}
\usepackage{nicefrac}
\usepackage{soul}
\usepackage{array}
\usepackage{siunitx}
\usepackage{bbold}
\usepackage[normalem]{ulem}
\usepackage[newcommands]{ragged2e}
\usepackage[dvipsnames]{xcolor}
\usepackage[font=small,labelfont=bf,justification=justified]{caption}
\usepackage{tikz}
\usetikzlibrary{shapes.geometric}

\usepackage{hyperref}
\hypersetup{
    colorlinks=true,
    citecolor=RedViolet,
    linkcolor=RedViolet,
    urlcolor=RedViolet
}

\newcounter{secnum}

\begin{document}

\title{Indefinite causal order with output-signalling instruments}

\author{Ricardo de Lima Silveira}
\thanks{These authors have equally contributed to this work}
\affiliation{Departamento de F\'isica Te\'orica e Experimental, Universidade Federal do Rio Grande do Norte, 59078-970 Natal-RN, Brazil}
\affiliation{International Institute of Physics, Federal University of Rio Grande do Norte, 59078-970, Natal, Brazil}
\author{Patrick Andriolo}
\thanks{These authors have equally contributed to this work}
\affiliation{Technische Universität Wien, Atominstitut \&  Vienna Center for Quantum Science and Technology, Stadionallee 2, 1020 Vienna, Austria}
\author{Davide Poderini}
\affiliation{Universit\`a degli Studi di Pavia, Dipartimento di Fisica, QUIT Group, via Bassi 6, 27100 Pavia, Italy}
\author{Dmitry Melnikov}
\affiliation{International Institute of Physics, Federal University of Rio Grande do Norte, 59078-970, Natal, Brazil}
\author{Rafael Chaves}
\affiliation{International Institute of Physics, Federal University of Rio Grande do Norte, 59078-970, Natal, Brazil}

\begin{abstract}
Quantum processes can exhibit indefinite causal order, allowing coherent superpositions of distinct orderings of events. Whether such processes generate genuinely noncausal statistics, however, depends not only on their global causal structure but also on the local operations available to the parties. Here, we restrict communication to the classical outcomes of local operations, yielding a causal polytope with new causal inequalities that can be violated using measure-and-reprepare, entanglement-breaking instruments requiring no coherent transmission through the laboratories. Using a semidefinite-programming framework, we bound the achievable violations and reveal noncausal correlations that satisfy all conventional input-signalling inequalities, demonstrating that restricting the information available for communication can uncover forms of indefinite causal order inaccessible under less restrictive operational assumptions.
\end{abstract}

\maketitle

\section{Introduction}

In recent years, nonclassical correlations have been explored in scenarios that extend or relax the assumptions underlying the paradigmatic Bell’s theorem \cite{Bell1964,Brunner2014}. Such generalizations include scenarios with multiple parties \cite{werner2001} or sources \cite{tavakoli2022bell}, as well as scenarios that allow communication between parties \cite{bacon2003bell,maxwell2014bell,chaves2017causal,chaves2015unifying,ringbauer2016experimental,brask2017bell,vieira2025can}. In the latter case, different operational restrictions on the information that can be communicated give rise to distinct classes of correlations, such as those in which parties can signal their inputs \cite{bacon2003bell,maxwell2014bell,chaves2017causal}, or their outputs \cite{chaves2015unifying,ringbauer2016experimental,brask2017bell,vieira2025can}. These operational assumptions determine the causal structure of the correlations from the outset. Remarkably, even this assumption of fixed causal structure can be relaxed: quantum processes can give rise to correlations that are incompatible with any well-defined causal order \cite{oreshkov2012,costa2026indefinite,rubino2017experimental,rozema2024experimental}. Such correlations are referred to as \textit{noncausal} correlations \cite{branciard2015simplest,miklin2017entropic,baumann2025no}.

Relaxing the assumption of a definite causal order is motivated both by fundamental and operational considerations. On the fundamental side, a consistent theory reconciling quantum theory with general relativity may require a framework in which causal order is not fixed \cite{hardy2005probability,hardy2007towards,hardy2018construction}. From an operational perspective, relaxing causal order extends the quantum circuit framework to processes in which the order of operations is not prescribed \cite{chiribella2013quantum,chiribella2016quantum}. A general framework for describing such processes was introduced in \cite{oreshkov2012}, where global processes are characterized by \textit{process matrices}, while the local operations available to the parties are described by \textit{instruments} \cite{Davies1970,wechs2021quantum}. The role of these two objects is depicted in Fig. \ref{fig: ICO scenario}. This framework gave rise to the notion of \textit{causal nonseparability}, which characterizes process matrices that cannot be decomposed into mixtures of process matrices with a definite causal order \cite{araujo2015witnessing}. When probed with suitable local instruments, such process matrices can generate probability distributions incompatible with any definite causal order \cite{branciard2015simplest,baumann2025no}, in a similar spirit, nonseparable states can generate nonlocal correlations, violating the well-known Bell inequalities \cite{Brunner2014}.

Here, we investigate a new indefinite causal order scenario that departs from previously considered constructions. Our new indefinite-causal-order scenario is based on the operational assumption that parties can communicate only the outputs of their local experiments. This is possible if the locally implementable instruments are restricted to measure-and-reprepare operations, which admit a direct realization in terms of POVMs and conditional state preparations. This particular set of entanglement-breaking instruments provide a simple operational setting for probing noncausal correlations, without requiring the preservation of quantum coherence through the laboratories \cite{ku2022quantifying,pal2015non,holevo2008entanglement,horodecki2003entanglement,shor2002additivity}. The restrictions defines a new causal polytope, characterized by 41 families of facet inequalities, 38 of which are nontrivial as compared to the only 2 nontrivial families of inequalities of the scenario with communication restriction considered in \cite{branciard2015simplest}.

Using semidefinite programming, we obtain reliable bounds on the violations achievable for each causal inequality. We show that output signalling can reveal noncausal correlations that remain undetected by the causal inequalities of the scenario of \cite{branciard2015simplest}. Our framework also accommodates imperfections in the local instruments, such as misaligned state preparations, allowing us to assess their impact on the observed violations.

The paper is organized as follows. Section~\ref{sec: related work} establishes the context for our results and compares them to the existing literature. In Sec.~\ref{sec: causal processes}, we introduce causal processes compatible with a definite causal order and review the input-signalling scenario of Ref.~\cite{branciard2015simplest} (Sec.~\ref{sec: input signalling}). We then introduce the output-signalling scenario considered in this work (Sec.~\ref{sec: output signalling}) and characterize its causal polytope. Section~\ref{sec: possibility of quantum violations} demonstrates that allowing quantum resources while preserving a definite causal order does not enlarge the corresponding set of output-signalling correlations, establishing that violations of its facet inequalities genuinely witness an indefinite causal order. In Sec.~\ref{sec: process matrix formalism}, we review the process-matrix formalism and the structure of the global processes used throughout the paper. Section~\ref{sec: see saw} details our numerical framework for optimizing causal-inequality violations using measure-and-reprepare instruments, with the resulting applications presented in Sec.~\ref{sec: numerical results}. Finally, we discuss our conclusions and open questions in Sec.~\ref{sec: conclusion}.

\begin{figure*}
    \centering
    \begin{minipage}[b]{0.475\textwidth}
        \centering
        \resizebox{1\linewidth}{!}{\input{images/ICO_scenario}}
        \caption*{\textbf{(a)} Setup of an indefinite causal order scenario.}
    \end{minipage}
    \hfill
    \hspace{0.5cm}
    \begin{minipage}[b]{0.475\textwidth}
        \centering
        \resizebox{1\linewidth}{!}{\input{images/measure_and_reprepare_channel}}
        \caption*{\textbf{(b)} Measure-and-reprepare instruments.}
    \end{minipage}
    \vspace{0.2cm}
    \caption{\textbf{(a)} Schematic representation of a scenario with global causal order between two distinct labs described by a \textit{process matrix} $W$. In their laboratories - where quantum theory is assumed to hold locally - Alice and Bob perform operations represented by instruments, denoted respectively by $\{M_{a\vert x}\}_{a,x} \in \mathcal{L}(\mathcal{H}^{A_I} \otimes \mathcal{H}^{A_O})$,  $\{M_{b\vert y}\}_{b,y} \in \mathcal{L}(\mathcal{H}^{B_I} \otimes \mathcal{H}^{B_O})$. The process matrix acts on $\mathcal{L} (\mathcal{H}^{A_I} \otimes \mathcal{H}^{A_O} \otimes \mathcal{H}^{B_I} \otimes \mathcal{H}^{B_O})$. \textbf{(b)} Anatomy of a measure-and-reprepare instrument $M^{A_I A_O}_{a\vert x}$. Here the system received by Alice is measured in a basis labelled by $x$ yielding a classical objective outcome $a$, and this pair is associated to a POVM element $E_{a\vert x}^{A_I} \in \mathcal{H}^{A_I}$. Based on the outcome $a$ registered by Alice, she prepares a quantum state $\rho_a^{A_O} \in \mathcal{D}(\mathcal{H}^{A_O})$, which is then sent again to Bob.}
    \label{fig: ICO scenario}
\end{figure*}

\section{Related work}
\label{sec: related work}

A central question in the study of indefinite causal order is how strongly its certification depends on the assumptions made about the operations implemented inside the laboratories \cite{rozema2024experimental}. Demonstrations of quantum control over indefinite causal order, such as implementations of the quantum switch, certified causal nonseparability through causal witnesses, a method that inherently required a detailed characterization of the implemented operations \cite{rubino2017experimental}. Other theoretical works have explored weaker assumptions on local devices by introducing semi-device-independent certification schemes, a setup in which only a subset of the local operations must be trusted \cite{bavaresco2019semi}. This approach demonstrated that the causal nonseparability of the quantum switch can be certified without fully characterizing all instruments. 

Alternatively, certification can be achieved by utilizing trusted quantum inputs while leaving the parties operations uncharacterized, or by imposing specific structures on the local instruments in a measurement-device-and-channel-independent scenario \cite{dourdent2022semi}. Fully device-independent certification of the quantum switch has also been shown to be possible; however, it requires an extended scenario involving an additional spacelike-separated party \cite{van2023device}, thus a distinct setup from the one considered in this work.

Our approach is complementary to these device- and semi-device-independent certification schemes. Rather than reducing the assumptions made about otherwise general local operations, here we investigate how physically restrictive these operations can be while still allowing for genuinely noncausal correlations. Every local instrument is constrained to an explicit measure-and-reprepare form: the incoming system is measured, its outcome is stored classically, and an outcome-dependent state is subsequently prepared. Because these corresponding channels are entanglement-breaking, they do not require the preservation of quantum coherence between the input and output of a laboratory. This constraint is both stronger than and conceptually distinct from previously considered structured instruments, which rely on trusted quantum inputs but leave the local measurements and channels uncharacterized \cite{dourdent2022semi}. 

While we restrict the implementable operations, the global process matrix is kept general, and we demonstrate that such elementary local operations are nevertheless sufficient to generate violations of causal inequalities within the output-signalling scenario.

On the experimental front, our results also address the ongoing challenge of implementing local measurements within processes exhibiting indefinite causal order. Implementations of a measurement operation inside a quantum switch required delaying the measurement readout to preserve coherence between the superimposed causal orders \cite{rubino2017experimental}. More recently, the aforementioned device-independent protocol \cite{van2023device} has been realized experimentally, demonstrating a violation of the corresponding causal inequality, albeit with remaining experimental loopholes \cite{richter2026toward}. Other works have emphasized the necessity of performing measurements with a local readout inside the quantum switch \cite{valibouse2026time}. In these setups, the readout is executed without destroying the signal's coherence by utilizing a time-delocalized auxiliary system, with the goal of evaluating a causal witness. Our results are thus of particular relevance, as they address the subtlety of extracting instrument statistics while simultaneously preserving an indefinite causal order.

\section{Causal Processes and Causal Inequalities}
\label{sec: causal processes}
Our goal is ultimately to study processes with indefinite causal order in a particular communication scenario. Before that, we must fully describe processes compatible with well-defined causal orders in that scenario. We will consider the following bipartite setting: the parties, A and B, are located in distinct semi-isolated laboratories, meaning that each one opens only once to receive or send out a physical system to the external world. Inside their laboratories, the experimentalists are able to prepare and to measure quantum states and to implement any quantum channel. Alice can choose between two operations, denoted by $x \in \{ 0,1 \}$, and each of these two local operations can result in one of two outcomes, denoted by $a \in \{0,1\}$. The same goes for Bob, with operation settings $y \in \{ 0,1 \}$ and possible outcomes $b \in \{0,1\}$. The correlations that the parties can establish are described by the joint probability distributions $p(a,b|x,y)$. They must satisfy nonnegativity conditions
    \begin{align}
        p(a,b|x,y)  \geq  0,  \quad \forall  \quad  a,b,x,y,
    \end{align}
    and normalization conditions for each pair of inputs
    \begin{align}\label{normalization}
        \sum_{a,b} p(a,b|x,y)  =  1, \quad \forall \quad   x,y.
    \end{align}

In a typical Bell scenario \cite{Bell1964, Brunner2014}, the parties are placed in a spacelike separation, so that no communication is possible between them. We can relax this condition and allow for scenarios with at most one-way communication, implemented in the following way: one of the parties receives a system, operates on it, and can then decide to encode some information on this system before sending it out to the other party.
    
In this protocol, the two parties can influence each other in two possible ways:
\begin{itemize}
    \item A acts first and is allowed to send information about her operations to B, but B cannot send any information to A. This situation is denoted by $A\prec B$.
    \item B acts first and is allowed to send information about her operations to A, but A cannot send any information to B. This situation is denoted by $B\prec A$.
\end{itemize}
In the first case, B's choice of operation cannot influence A's outcome, so the correlations $p^{A\prec B}(a,b|x,y)$ must satisfy no-signalling-to-A conditions:
\begin{align}\label{nonsig_A}
         \sum_b p^{A \prec B}(a,b|x,y) =  \sum_b p^{A \prec B}(a,b|x,y^{\prime}) \quad \forall  x,y,y',a.    
\end{align}
Analogously, the correlations $p^{B\prec A}(a,b|x,y)$ must satisfy no-signalling-to-B conditions:
\begin{align}\label{nonsig_B}
        \sum_a p^{B \prec A}(a,b|x,y) =  \sum_a p^{B \prec A}(a,b|x^{\prime},y) \quad \forall  x,x',y,b.
\end{align}

The most general probability distribution compatible with a well-defined order of events is given by a statistical mixture of the two cases above, that is:
\begin{align}\label{general_causal}
    p(a,b|x,y) = q  p^{A\prec B}(a,b|x,y) + (1-q)  p^{B\prec A}(a,b|x,y), 
\end{align}
for $q\in [0,1]$. To make sense of this equation, we can think of a multi-round experiment in which the order for the operations of A and B is determined probabilistically before each round, so the order is not fixed beforehand but is well-defined for each round.

The complete statistical description of the experiment is captured by the set $\mathcal{P} = \{ p(a,b|x,y) \}_{a,b,x,y \in \{0,1\}}$, referred to as a \textit{process}. Mathematically, each process can be represented as a vector within a 16-dimensional probability space. A process compatible with Eq. (\ref{general_causal}) is defined as a \textit{causal process}; otherwise, it is said to exhibit an \textit{indefinite causal order} (ICO) \cite{Oreshkov2016}. In a classical description, the set of all bipartite processes with a definite causal order forms a convex polytope, meaning it can be completely characterized by a finite number of linear causal inequalities. In this work, we derive classical causal inequalities for a particular communication scenario in Sec. \ref{sec: output signalling} and show in Sec. \ref{sec: possibility of quantum violations} that these inequalities remain valid for quantum processes that possess a definite causal order. Consequently, any violation of these inequalities genuinely witnesses an indefinite causal order, irrespective of the underlying classical or quantum nature of the correlations.
    
\subsection{Input signalling}
\label{sec: input signalling}

A setup that implements one particular type of one-way communication was studied in Ref. \cite{branciard2015simplest}. There, it was shown that the sets of processes compatible with either $A\prec B$ or $B\prec A$ are both convex polytopes, and the convex hull of these two contains all the processes with correlations that satisfy Eq. (\ref{general_causal}). Here we briefly review their approach and key results.

By employing the one-way communication conditions and Bayes' rule, the correlations can be expanded in terms of deterministic distributions in the following way:
\begin{subequations}\label{inp_sing_expan}
    \begin{align}
            p^{A \prec B}(a,b|x,y) \      
          &= \ \sum_{\alpha,\beta}  q_{\alpha,\beta} \ \delta_{a,\alpha(x)} \ \delta_{b,\beta(x,y)} \\
            p^{B \prec A}(a,b|x,y) \      
         &= \ \sum_{\alpha,\beta}  q_{\alpha,\beta}^{\prime} \ \delta_{a,\alpha(x,y)} \ \delta_{b,\beta(y)},
    \end{align}        
\end{subequations}
with $q_{\alpha,\beta}^{(\prime)} \ge 0, \ \sum_{\alpha,\beta} q_{\alpha,\beta}^{(\prime)}   = 1$. The sum is taken over all possible deterministic functions from inputs to outputs. Then, correlations of the form of Eq. (\ref{general_causal}) form a convex set whose extremal points are given by these deterministic distributions. This set is the so-called \textit{causal polytope}. 
    
Upon characterizing the causal polytope, Ref. \cite{branciard2015simplest} found 48 facets, 16 corresponding to nonnegativity conditions and 32 associated with nontrivial \textit{causal inequalities}. The latter can be divided into two families of inequalities equivalent under relabelings of inputs and outputs: 16 are equivalent to the \textit{guess your neighbor's input} (GYNI) inequality
\begin{equation}\label{gyni}
        \mathcal{I}_\text{GYNI}=\sum_{x,y,a,b} \delta_{a,y}  \delta_{b,x} p(a,b|x,y)\; \leq \; 2,
\end{equation}
and the other 16 are equivalent to the \textit{lazy guess your neighbor's input} (LGYNI) inequality
     \begin{equation}\label{lgyni}
        \mathcal{I}_\text{LGYNI} = \sum_{x,y,a,b} \delta_{x(a \oplus y),0}  \delta_{y(b \oplus x),0} p(a,b|x,y)\; \leq \; 3.
    \end{equation}
    
Although no explicit restriction was imposed in Ref. \cite{branciard2015simplest} on what type of communication is allowed, one sees from Eqs.(\ref{inp_sing_expan}) that these sets of correlations are equivalent to those that can be achieved when only the input of one party can directly influence the other party's outcome. This \textit{input signalling} is represented by the directed acyclic graphs (DAGs) in Fig. \ref{fig:signalling}. Following the Markov decomposition \cite{chaves2015unifying,pearl2009causality}, the causal influences captured by these DAGs are given by the following models of hidden variables: If  $A \prec B$, then the causal model assumes the decomposition
 \begin{align}
     p^{X \rightarrow B}(a,b|x,y) &= \sum_{\lambda} \; p(\lambda)p(a|x,\lambda)p(b|x,y,\lambda).  \label{inp_sing_fromdag_a}
 \end{align}
Similarly, when B preceeds A ($B \prec A$), then we have
 \begin{align}
 \quad p^{Y\rightarrow A}(a,b|x,y) &= \sum_{\lambda} \; p(\lambda)p(b|y,\lambda)p(a|x,y,\lambda). \label{inp_sing_fromdag_b}
 \end{align}

\begin{figure}
    \centering
    \begin{minipage}[b]{0.476\columnwidth}
        \centering
        \resizebox{1\linewidth}{!}{\input{images/input_signalling}}
        \caption*{\textbf{(a)} Signalling of inputs.\label{fig:inp_sing_dags}}
    \end{minipage}
    \hfill
    \begin{minipage}[b]{0.46\columnwidth}
        \centering
        \resizebox{1\linewidth}{!}{\input{images/output_signalling}}
        \caption*{\textbf{(b)} Signalling of outputs. \label{fig:out_sing_dags}}
    \end{minipage}
    \caption{DAGs representing scenarios with communication between Alice and Bob, with a common source $\Lambda$. In \textbf{(a)} the communication regards the possible inputs of the parties, while \textbf{(b)} shows communication of their outputs.}
    \label{fig:signalling}
\end{figure}

 In Eq.(\ref{inp_sing_fromdag_a}), each single party probability distribution can be expanded in terms of deterministic distributions: $p(a|x,\lambda) = \delta_{a,f_{\lambda}(x)}$ and $p(b|x,y,\lambda) = \delta_{a,g_{\lambda}(x,y) }$. Summing over all $\lambda$ is equivalent to considering all possible deterministic functions, so that Eq.(\ref{inp_sing_expan}) is recovered. The same reasoning applies to Eq.(\ref{inp_sing_fromdag_b}).
 
\subsection{Output signalling}
\label{sec: output signalling}

We now consider a more restrictive scenario by explicitly imposing a constraint on the type of information that can be communicated by each party:  after performing their local operations, A and B can only send out information about the value of their outcomes, but not about their inputs \cite{pawlowski2010non,ringbauer2016experimental,brask2017bell}.
        
Again, we first characterize the situations in which the parties operate in well-defined orders: they can either precede one another or be space-like separated, in which case there can be no communication. If A precedes B $(A \prec B)$, B cannot send any information to A, so the no-signalling-to-A conditions of Eq. (\ref{nonsig_A}) must hold. Similarly, when B precedes A $(B \prec A)$, the no-signalling-to-B conditions of Eq. (\ref{nonsig_B}) must be satisfied. If the parties cannot communicate, both no-signalling conditions are satisfied.

In the $A \prec B$ case, the key difference in comparison to the input-signalling scenario is that B's outcome $b$ can no longer depend directly on A's measurement setting $x$. The outcome $b$ will explicitly depend only on $a$ and $y$: 
\begin{align}\label{out_sing_fromdag_a}
    p^{A\rightarrow B}(a,b|x,y) = \sum_{\lambda} \; p(\lambda)p(a|x,\lambda)p(b|y, a,\lambda).
\end{align}
Similarly, in the $B \prec A$ case $a$ can only depend on $b$ and $x$. Thus, the correlations can be written in the following manner:
\begin{align}\label{out_sing_fromdag_b}
    p^{B\rightarrow A}(a,b|x,y) = \sum_{\lambda} \; p(\lambda)p(b|y,\lambda)p(a|x,b,\lambda).
\end{align}

The graphical representation of this expression is provided by the DAGs in Fig \ref{fig:signalling}, and by expanding the correlations of each party in terms of deterministic distributions, we obtain probabilities with the following decomposition:

\begin{subequations}\label{out_sing_expan}
    \begin{align}\label{out_sing_expan_a}
        p^{A\rightarrow B}(a,b|x,y) \      
      &= \ \sum_{\alpha,\beta}  q_{\alpha,\beta} \ \delta_{a,\alpha(x)} \ \delta_{b,\beta(a,y)}, \\ 
        p^{B \rightarrow A}(a,b|x,y) \      
      &= \ \sum_{\alpha,\beta}  q_{\alpha,\beta}^{\prime} \ \delta_{a,\alpha(x,b)} \ \delta_{b,\beta(y)} ,\label{out_sing_expan_b}
    \end{align}        
\end{subequations}
with $q_{\alpha,\beta}^{(\prime)} \ge 0, \ \sum_{\alpha,\beta} q_{\alpha,\beta}^{(\prime)}   = 1.$     

From Eq. (\ref{out_sing_expan_a}), the set of processes compatible with $A \prec B$ is a convex combination of a finite number of points in the probability  space. This set is therefore a polytope, whose vertices are the deterministic causal processes with entries given by deterministic distributions. To obtain these vertices, one simply needs to list all possible combinations of the deterministic functions $\alpha(x)$ and $\beta(a,y)$. The same procedure is carried out for the set of processes compatible with $B\prec A$ and described by Eq. (\ref{out_sing_expan_b}). We find that each of these polytopes has 40 extremal points. Together, they have 64 unique (non repeated) extremal points, the convex hull of which is the causal polytope.

Using the LRS software \cite{avis1991pivoting} for facet enumeration, we found that the output-signalling causal polytope comprises 276 facets. Of these, 16 correspond to trivial nonnegativity constraints, $p(a,b|x,y) \geq 0$, while the remaining 260 define nontrivial causal inequalities.

To classify these inequalities into equivalence families, recall that in standard Bell scenarios and conventional input-signalling setups, equivalent inequalities are generated via discrete symmetry operations: relabeling parties, inputs, or outputs conditioned on inputs \cite{Rosset2014, Scarani2019}. In the output-signalling scenario, however, output relabelings cannot be conditioned on inputs, as doing so would violate the core assumption that no input information is communicated.

Applying this restricted set of valid relabelings partitions the 260 nontrivial causal inequalities into 38 distinct equivalence families, cataloged in Appendix \ref{app: inequality list}. Interestingly, the GYNI inequality is no longer a facet-defining causal inequality in the output-signalling framework. Conversely, while several symmetries of the LGYNI inequality remain facet-defining, not all symmetries valid in the input-signalling case persist here due to the restriction on input-dependent relabelings.

\subsection{Absence of classical-quantum separation}
\label{sec: possibility of quantum violations}

We note that the input signalling DAGs in Fig. \ref{fig:signalling} and the corresponding equations constitute a classical hidden variable model. But can quantum correlations with definite causal order violate the corresponding causal inequalities? In other words, are the quantum sets characterized by behaviours of the form
\begin{subequations}\label{quant_inp_sing}
        \begin{align}
        p^{A \rightarrow B}_{\mathcal{Q}}(a,b|x,y) &= \text{Tr}\left( A^{a}_{x} \otimes B^{b}_{x,y}  \rho \right), \\
        p^{B \rightarrow A}_{\mathcal{Q}}(a,b|x,y) &= \text{Tr}\left( A^{a}_{x,y} \otimes B^{b}_{y}  \rho \right),
    \end{align}
\end{subequations}
larger than the sets described by Equations (\ref{inp_sing_fromdag_a}) and (\ref{inp_sing_fromdag_b})? In the input-signalling case, the answer is no, as the literature has established that this type of communication is sufficient to simulate quantum correlations \cite{bacon2003bell,chaves2017causal}. Consider the case $A \prec B$. Any correlation that satisfies no-signalling-to-A (including quantum correlations) can be written in the form
\begin{equation}\label{NS_b_to_a}
    p(a,b|x,y) = p(b|x,y,a)  p(a|x).
\end{equation}

Examining Eq. (\ref{inp_sing_fromdag_a}), we see that $p(b|x,y,a)$ can always be generated locally by B, given that he has access to $x$ and $y$ (for each $\lambda$, $a$ is determined deterministically by $x$), so the set of correlations of the type of Eq. (\ref{NS_b_to_a}) can be completely generated with input communication. There is no classical-quantum gap in this case. An analogous argument applies to the $B \prec A$ order of events.

Now, for the output signalling scenario we are investigating, the argument above does not apply directly. For $A \prec B$, it is not guaranteed that the no-signalling-to-A condition in Eq. (\ref{NS_b_to_a}) can always be simulated if one has access to correlations of the type given in Eq. (\ref{out_sing_fromdag_a}). The quantity $p(b|x,y,a)$ cannot always be locally generated by knowing $p(b|a,y,\lambda)$, since knowledge of $a$ and $\lambda$ might not be enough to always determine $x$. In principle, the quantum sets defined by
\begin{align}
    p^{A \rightarrow B}_{\mathcal{Q}}(a,b|x,y) &= \text{Tr}( A^{a}_{x} \otimes B^{b}_{a,y}  ), \\
    p^{B \rightarrow A}_{\mathcal{Q}}(a,b|x,y) &= \text{Tr}( A^{a}_{x,b} \otimes B^{b}_{y})
\end{align}
could be larger than their classical counterparts described in Equations (\ref{out_sing_fromdag_a}) and (\ref{out_sing_fromdag_b}). If so, some quantum processes compatible with well-defined causal structures could violate causal inequalities of the output signalling polytope, rendering those inequalities illegitimate witnesses of ICO.

However, using tools from graph theory, we show that for the simplest output signalling scenario with binary variables, the classical and quantum sets compatible with definite causal order are indeed the same. The argument is based on the Cabello-Severini-Winter (CSW) graph theoretical framework, originally introduced to investigate contextuality \cite{Cabello2014} and later extended to causal modeling \cite{pmlr-v115-poderini20a}. The core idea is to encode the relevant causal relations in an undirected graph called the \textit{exclusivity graph}. We can then obtain information about the sets of classical and quantum correlations attainable in a given scenario from properties of the corresponding exclusivity graph.

Let us consider the case in which A performs $x$ and obtains $a$, while B performs $y$ and obtains $b$. This constitutes an \textit{event} denoted by $e \; = a,b| x,y$. Two events $e$ and $e^{\prime}$ are considered to be exclusive if they cannot happen together given a particular choice of experimental setting. This happens if there is a measurement process able distinguish between the outcomes, so that they cannot happen simultaneously in the same run of the experiment. 

In a dichotomic Bell scenario, for instance, where the distribution $p(ab|xy)=\text{Tr}(\rho \cdot E_{a \vert x} \otimes E_{b\vert y})$ is associated to events where one can obtain $a,b\in \{0,1\}$, Alice obtains an exclusive events if she obtains answers $a=0$ and $a=1$ while setting $x = 0$ for both, as only one of them can happen in a single run of the experiment. On the contrary, if the setting $x$ is different (for example $x = 1$ for $a=0$ and $x = 0$ for $a=1$), the events will not be exclusive since a single experimental test cannot distinguish between them.

In general, a graph is an ordered pair $G = (V,E)$, where $V$ is a set of vertices and $E$ is a set of edges, that is, pairs $(u,v)$, $u,v \in V$. The exclusivity graph of a given scenario is the one whose vertices are the events in that scenario. Two vertices in this graph are connected by an edge if and only if they correspond to exclusive events.

Each event $a,b|x,y$ in the exclusivity graph is associated to the probability distribution $p(a,b|x,y)$. These distributions may form different sets depending on what physical theory generates the statistics. A typical task is to distinguish between sets of classical and nonclassical correlations. The classical correlations are bounded by hyperplanes associated with linear inequalities which can be used to probe different manifestations of nonclassicality, such as nonlocality \cite{Brunner2014,rabelo2014multigraph} and contextuality \cite{amaral2018graph,amaral2017geometrical}.

Let us consider a set of events $\{e_{i}\}$ with exclusivity graph $G$. A typical linear inequality witnessing nonclassicality for these events has the form $\mathcal{I} = \sum_{i} w_{i} p_{i} \leq \mathcal{I_{\beta}}$. All classical correlations must satisfy any such inequality, whereas some nonclassical correlations may violate it.

The expression $\mathcal{I}$ can be studied with the aid of the exclusivity graph if we work not just with $G$, but append to it the coefficiets $w_{i}$ as weigths to the nodes of the graph. The resulting object is the vertex-weigthed graph denoted by $(G, w)$. $\mathcal{I} $ can then be seen as a function of the graph $G$ and of the set of weights $w$, $\mathcal{I} = \mathcal{I}(G, w)$

The CSW framework provides the means to calculate the classical and quantum bounds of $\mathcal{I}(G, w)$ from properties of the weighted exclusivity graph. For a classical realization of $G$, $\mathcal{I}(G, w)$  is bounded by the \textit{independence number} of $G$, denoted by $\alpha(G,w)$, while the maximum quantum value is upper-bounded by the Lovász number, $\vartheta(G,w)$. These results can be summarized in the expression
\begin{align}
    \mathcal{I}(G, w) \; \overset{\mathcal{C}}{\leq} \;  \alpha(G,w) \;\overset{\mathcal{Q}}{\leq} \; \vartheta(G,w). 
\end{align}

In some cases, the classical and quantum realizations of $G$ are actually the same set, and $\alpha(G,w) = \vartheta(G,w)$. That happens when the graph is of a special type called \textit{perfect}. From the \textit{strong perfect graph theorem}, a graph is perfect if it has no induced $n$-cycle, with $n$ odd and $n  \geq  5$ \cite{Chudnovsky2006}.

To see how these results can be applied to our problem, let us consider the case $A \prec B $. We can see that two events $a^{},b^{}  |  x^{},y{}$ and $a^{\prime},b^{\prime}  |  x^{\prime},y{\prime}$ are exclusive if at least one of the following conditions holds: (i) $x=x^{\prime}$ and $a \neq a^{\prime}$; (ii) $y=y^{\prime}$, $a=a^{\prime}$ and $b \neq b^{\prime}$.  With this in mind, we build the exclusivity graph shown in Fig.(\ref{fig:A_to_B_exc_graph})

\input{images/A_to_B_exc_graph}

One can directly check that it matches the criteria presented above for a graph to be considered perfect and therefore the set of correlations $p^{A \rightarrow B}(a,b|x,y)$ that can be achieved in this scenario is the same whether we use shared classical randomness or perform measurements on a quantum state.
The same reasoning applies to the $B \prec A$ case. 

Thus, even though the causal polytope was constructed with classical random variables, it also contains all quantum correlations compatible with definite causal order. Thus, a violation of a causal inequality of the output signalling polytope cannot be simply obtained by local measurements on shared entangled states; it is a legitimate witness of indefinite causal order.

\section{Process Matrix Formalism}
\label{sec: process matrix formalism}

We have described the causal correlations in a bipartite scenario where we impose constraints on the type of information a party can share with others. We now investigate the possibility of generating correlations stronger than those contained in the causal polytope; that is, we want to study the possibility of violating our causal inequalities. 

To that end, we employ the \textit{process matrix formalism} introduced in Ref. \cite{oreshkov2012}, a generalization of standard quantum mechanics that still accounts for processes occurring in a well-defined causal order, be it fixed or dynamic, but also allows for processes that are not compatible with any definite causal order.

The fundamental goal is to construct a framework that preserves the quantum mechanical descriptions of the operations in each laboratory, but that does not specify \textit{a priori} a global causal structure on which the parties are embedded. 

The laboratory of a given party, say A, is associated with an input Hilbert space $\mathcal{H}^{A_{I}}$ and an output Hilbert space $\mathcal{H}^{A_{O}}$. These Hilbert spaces have dimensions $d_{A_{I}}$ and $d_{A_{O}}$ respectively, but we can take them to be equal without loss of generality: $d_{A_{I}}  =  d_{A_{O}}  \equiv  d_{A_{}}$. A quantum system prepared in a state $\rho^{A_{I}} \in \mathcal{L}(\mathcal{H}^{A_{I}})$ can enter the laboratory, where $\mathcal{L}(\mathcal{H}^{A_{I}})$ denotes the space of linear operators over $\mathcal{H}^{A_{I}}$. A can then perform measurements on that system or evolve it through some quantum channel, and can also reprepare this system in a state $\rho^{A_{O}} \in \mathcal{L}(\mathcal{H}^{A_{O}})$ to be sent out of the laboratory. A unified description of these local operations is provided by the concept of a \textit{quantum instrument} \cite{Davies1970}: a set of completely positive (CP) maps from $\mathcal{L}(\mathcal{H}^{A_{I}})$ to $\mathcal{L}(\mathcal{H}^{A_{O}})$, each associated with one possible outcome of the local operation, with the requirement that their sum results in a CPTP map. 
    
The Choi-Jamiolkowski (CJ) isomorphism \cite{Choi1975, Jamiokowski1972} provides a convenient matrix representation of these maps. The complete positivity and trace-preserving conditions translate to positive semidefinite and completeness conditions. The situation in which Alice performs the operations associated with input $x$ and obtains the outcome labeled by $a$ is described by the CJ matrix $M_{a|x}^A \in \mathcal{L}(\mathcal{H}^{A_{I}} \otimes \mathcal{H}^{A_{O}})$. A quantum instrument is then given by a set $\{M_{a|x}^A\}_{a=0}^{m-1}$, satisfying 
\begin{equation}\label{quantum_inst_conds}
   M_{a|x}^{A_IA_O} \geq 0 \ \forall \  a  \quad \mathrm{and} \quad \text{Tr}_{A_O} \sum_a M_{a|x}^{A_IA_O} = \mathbb{1}^{A_I}.
\end{equation}
    
If the probabilities that parties obtain the outcomes $a$ and $b$ after choosing operations $x$ and $y$ are to be compatible with an ordinary quantum mechanical description in each laboratory, they must have the form \cite{oreshkov2012,araujo2015, gleason1975measures}
    \begin{equation}\label{gen_born}
       p(a,b|x,y) = \text{Tr}\big[ (M_{a|x}^{A_IA_O} \otimes M_{b|y}^{B_IB_O}) \cdot W\big],
    \end{equation}
where $W \in \mathcal{L}(\mathcal{H}^{A_I} \otimes \mathcal{H}^{A_O} \otimes \mathcal{H}^{B_I} \otimes \mathcal{H}^{B_O})$. The requirement that the quantities in Eq. (\ref{gen_born}) be well-defined probability distributions imposes some restrictions on the object $W$, namely:
    \begin{subequations}\label{eq:valid_W}
    \begin{gather}
     W \ge 0  , \\[1mm]
     \text{Tr}(W) = d_{A_O}  d_{B_O}  , \\[1mm]
     {}_{B_IB_O}W = {}_{A_OB_IB_O}W  , \\[1mm]
     {}_{A_IA_O}W = {}_{A_IA_OB_O}W  , \\[1mm]
     W = {}_{B_O}W + {}_{A_O}W - {}_{A_OB_O}W  ,
    \end{gather}
    \end{subequations}
    where the trace-and-replace operation $_X \cdot$ is defined as
    \begin{equation}
        _X W = \frac{\mathbb{1}^{X}}{d_X} \otimes \text{Tr}_X W.
    \end{equation}

If these conditions are satisfied, $W$ is called the \textit{process matrix} and can be interpreted as encoding the possible causal connections between laboratories. More precisely, considering a basis $\{\sigma_\mu^{X}\}_{\mu=0}^{d^2_X - 1}$ for the space $\mathcal{L}(\mathcal{H}^{X})$, a general decomposition of $W$ with the form
    \begin{align}\label{eq: W basis decomposition}
W=\sum_{\alpha,\beta,\gamma,\delta=0}^{3}
w_{\alpha\beta\gamma\delta}\,
\sigma_{\alpha}^{A_I}\otimes
\sigma_{\beta}^{A_O}\otimes
\sigma_{\gamma}^{B_I}\otimes
\sigma_{\delta}^{B_O},
\end{align}
can admit only certain terms, each representing signalling compatible with certain causal directions. For instance, terms compatible solely with $A \prec B$ are supported by strings with the structure \cite{costa2026indefinite}
\begin{align}
\sigma_\alpha^{A_I}\sigma_\beta^{A_O}
 \sigma_\gamma^{B_I}\mathbb{1}^{B_O},
 \qquad \beta,\gamma\neq0,\label{eq:class-2}
\end{align}
while terms related exclusively to $B \prec A$ are structured according to
\begin{align}
   \sigma_\alpha^{A_I}\mathbb{1}^{A_O}
\sigma_\gamma^{B_I}\sigma_\delta^{B_O},
 \qquad \alpha,\delta\neq0.\label{eq:class-3}
\end{align}
It's possible to have terms compatible with both $A \prec B$ and $B \prec A$. Those assume the form
\begin{align}
\sigma_\alpha^{A_I}\mathbb{1}^{A_O}
\sigma_\gamma^{B_I}\mathbb{1}^{B_O}.\label{eq:class1}
\end{align}
Assuming interpretations in terms of causal directions, terms of the form (\ref{eq:class-2})-(\ref{eq:class1}) can  be seen as descriptions of states and channels. These are the following \cite{oreshkov2012}:
\begin{itemize}
    \item[\textit{1)}] \textit{States} correspond to terms with only $A_I$ \textit{and/or} $B_I$ having nontrivial information, as in Eq. (\ref{eq:class1});

    \item[\textit{2)}] \textit{Channels} are related to terms with non-identity terms in $A_O B_I$ (a channel connecting Alice's output to Bob's input, $A \prec B$) or $A_I B_O$ (an output from Bob going to Alice's lab $B \prec A$);
\end{itemize}

\begin{itemize}
    \item[\textit{3)}] \textit{Channels with memory} are indicated by terms where one party has full information in its input and output system, while the other has only input information, i.e. $A_I A_O B_I$ when both input and output from Alice influence Bob's input ($A \prec B$) and vice-versa. These channels combine the two previous options, as in the case of an entangled state shared by $A$ and $B$, suplemented by a one-way channel.
\end{itemize}

Other terms, which might represent local or global loops (e.g. $A_I A_O$ or $A_I A_O B_O$, respectively) are forbidden by the process matrix constraints, as well as postselection terms ($A_O$, $B_O$, $A_O B_O$) as they don't respect normalization.

\section{An optimization framework for violating output-signalling causal inequalities with measure-and-reprepare instruments}\label{sec: see saw}

We now apply the process matrix formalism to describe scenarios where parties can only communicate the classical outcomes of their local interventions. To this end, we restrict local operations to measure-and-reprepare procedures: upon receiving a quantum system, a laboratory performs a measurement and subsequently prepares and transmits a new state conditioned on the observed outcome. For party $A$, the CJ operator describing this measure-and-reprepare instrument is given by
\begin{equation}\label{measure_reprep_cj}
  M^{A_{I}A_{O}}_{a|x} = E^{A_{I}}_{a|x} \otimes (\rho^{T} _{a})^{A_{O}},
\end{equation}
where $E^{A_{I}}_{a|x}$ is a POVM element measured in $A_{I}$ and $\rho _{a}$ is a state prepared in $A_{O}$ encoding information about the output $a$. Naturally, these objects must satisfy the standard conditions imposed on POVMs and states:
    \begin{align}
     &E^{A_{I}}_{a|x} \; \geq \; 0; \quad  \quad \sum_{a}E^{A_{I}}_{a|x} = \mathbb{1}^{A_I} &\forall  a, x, 
    \end{align}
and $\rho_{a}  \geq  0  , \quad \text{Tr}(\rho _{a}) = 1$ for all $a$. With this, one can easily verify that the set of $M^{A_{I}A_{O}}_{a|x}$ satisfies the conditions in Eq. (\ref{quantum_inst_conds}) for well-defined quantum instruments. The description of $M^{B_{I}B_{O}}_{b|y}$ is analogous. We recall that we are interested in a scenario where each party can choose between two dichotomic measurements, so $a,b,x,y \in {0,1}$. With that in mind, we work with two-dimensional systems, $\mathrm{dim} (\rho_a) = \mathrm{dim}  (\rho_b) = 2$. 

In general, a causal inequality is a linear combination of probability distributions bounded by an optimal value, $\beta_\mathcal{C}$, achievable within causal process matrices:
    \begin{equation}
        \mathcal{I}  = \sum_{a,b,x,y} c _{abxy}  p(a,b|x,y) \; \geq \; \beta_\mathcal{C}.
    \end{equation}
Let $\mathcal{E_{A}} = \{ E^{A_{I}}_{a|x} \}_{x,a}$ and $\mathcal{E_{B}} = \{ E^{B_{I}}_{b|y} \}_{y,b}$ be the POVMs of A and B, respectively, and $\mathcal{D_A} = \{ \rho_{a}\}_{a}$ and $\mathcal{D_B} = \{ \rho_{b}\}_{b}$ the sets of states they prepare. From Eqs. (\ref{gen_born}) and (\ref{measure_reprep_cj}), $\mathcal{I}$ is of the form $\mathcal{I}(\mathcal{E_A}, \mathcal{D_A}, \mathcal{E_B}, \mathcal{D_B}, W)$. To find violations of the inequality, we want to optimize this expression as a function of positive semidefinite operators. 

The usual approach for finding bounds on objective functions which are polynomials on a certain number of operators typically relies on using semi-definite relaxations, such as the paradigmatic Navascues-Pironio-Acin (NPA) hierarchy \cite{npa2008,npa2007}. In our case, however, the constraints defining the set of valid process matrices inherently depend on the tensor product structure of the theory, while SDP relaxations for noncommutative polynomials typically relax tensor compositions to commutation relations. Thus, SDP relaxations are not suitable for the problem posed in the maximization of a causal inequality. 

In order to circumvent the impossibility of translating the problem of minimizing a causal inequality with SDP relaxations, we perform the optimization iteratively, implementing a \textit{see-saw} algorithm  \cite{werner2001, branciard2015simplest,Abbott2016}. This allows us to find reliable bounds on the maximal violation of such inequalities. At each step, we minimize the objective function of a desired outcome-signalling causal inequality over a subset of its arguments, in such a way that the problem can be solved with standard SDP techniques. 

As a first step, we generate random sets of POVM's and output states for Alice and Bob: $\mathcal{E}^{({0})}_\mathcal{A} $, $\mathcal{E}_\mathcal{B}^{({0})} $, $\mathcal{D}_\mathcal{B}^{({0})} $ and $\mathcal{D}_\mathcal{B}^{({0})} $. These initially remain fixed while the expression is minimized over the process matrices:
\begin{subequations}\label{eq:w_sdp}
    \begin{align}
    \min_W \quad &\mathcal{I}(\mathcal{E}^{(0)}_\mathcal{A}, \mathcal{D}^{(0)}_\mathcal{A}, \mathcal{E}_\mathcal{B}^{(0)}, \mathcal{D}_\mathcal{B}^{(0)}, W)\\
    \text{s.t.} \quad &W \ge 0  , \\[1mm]
     &\text{Tr}(W) = d_{A_O}  d_{B_O}  , \\[1mm]
     &{}_{B_IB_O}W = {}_{A_OB_IB_O}W  , \\[1mm]
     &{}_{A_IA_O}W = {}_{A_IA_OB_O}W  , \\[1mm]
     &W = {}_{B_O}W + {}_{A_O}W - {}_{A_OB_O}W  .
    \end{align}
\end{subequations}

This results in an optimal process matrix $W^{(1)}$ that is then fixed while optimizations are performed over instruments. To find each instrument according to the measure-and-reprepare strategies specified in Eq. (\ref{measure_reprep_cj}), we perform again a see-saw, this time between POVM's and density matrices for both parties. For Alice, the optimal set of POVM's is found with
\begin{subequations}\label{eq:e_A_sdp}
    \begin{align}
    \min_{\{\mathcal{E}_\mathcal{A}\}} \quad &\mathcal{I}(\mathcal{E}_\mathcal{A}, \mathcal{D}_\mathcal{A}^{(0)}, \mathcal{E}_\mathcal{B}^{(0)}, \mathcal{D}_\mathcal{B}^{(0)}, W^{(1)})\\
    \text{s.t.} \quad &E^{A_{I}}_{a|x} \geq  0 \quad  &\forall \  a,x, \\
    &\sum_{a}E^{A_{I}}_{a|x} = \mathbb{1}^{A_I} \quad  &\forall  \ x.
\end{align}
\end{subequations}
The set of POVM's that realize the optimal value of the inequality in the optimization above then replaces the initial random measurements, $\mathcal{E}^{(0)}_\mathcal{A} \mapsto \mathcal{E}^{(1)}_\mathcal{A}$. After this, another SDP problem searchs for the optimal output states of Alice,
    $\mathcal{D}_\mathcal{A}^{(1)}$:
    \begin{subequations}\label{eq:r_A_sdp}
    \begin{align}
    \min_{\{\mathcal{D}_\mathcal{A}\}} \quad &\mathcal{I}(\mathcal{E}_\mathcal{A}^{(1)}, \mathcal{D_A}_{}, \mathcal{E}_\mathcal{B}^{(0)}, \mathcal{D}_\mathcal{B}^{(0)}, W^{(1)}) \\
    \text{s.t.}\quad &\rho_{a} \geq 0 , \quad \text{Tr}(\rho_{a}) = 1 &\forall \  a.
    \end{align}
    \end{subequations}
Analogous SDP minimizations are performed for Bob's POVMs and states, completing a cycle of five optimization steps. The cycles are repeated $n$ times until the inequality expression converges to a reliable upper bound $\mathcal{I}(\mathcal{E}^{({n})}_\mathcal{A} , \mathcal{E}_\mathcal{B}^{({n})} , \mathcal{D}_\mathcal{B}^{({n})},\mathcal{D}_\mathcal{B}^{({n})}, W^{(n)})$ for the global minimum value achievable within process matrix correlations and measure-and-reprepare instruments with fixed dimensions. The algorithm used for the see-saw can be found in \cite{OutputSignallingData}.

\section{Violation of causal inequalities with measure-and-reprepare instruments}
\label{sec: numerical results}

In this section, we employ the optimization framework presented in Sec. \ref{sec: see saw} to showcase how noncausal correlations appear in the scenario with signalling of outputs via measure-and-reprepare instruments.

Our results are organized as follows. In section \ref{sec: comparing violations} we start by examining, from the output signalling perspective, a noncausal process matrix previously studied in the scenario with signalling of inputs \cite{branciard2015simplest}. We provide the optimal violation that this process matrix can achieve within our scenario, and we also exhibit a simple set of states and POVM's that can be employed by the agents to extract noncausal statistics from this process. In Sec. \ref{sec: process matrix tailored for output signalling} we constraint our optimization method to correlations that do violate output-signalling inequalities while respecting the input-signalling facets, providing examples of correlations that live in the gap of input-signalling and output-signalling strategies. Furthermore, we show in Sec. \ref{sec: orthogonality thresholds} that overlaps between output states hinder noncausality in the output signalling scenario, indicating that orthogonal states appear to be optimal. Finally, in Sec. \ref{sec: violation bounds} we display the optimal violations found for each class of inequalities found in the output signalling scenario.

\subsection{Comparison with the input signalling scenario}
\label{sec: comparing violations}

As a first illustration of the output signalling strategy, we provide a comparison of the violation of the LGYNI inequality in the output-signalling scenario considered in this work and in the input-signalling case of Ref. \cite{branciard2015simplest}. 

In order to directly compare $\mathcal{I}_\text{GYNI}$ of Eq. (\ref{gyni}) and $\mathcal{I}_\text{LGYNI}$ of Eq. (\ref{lgyni}) to the inequalities derived in the output-signalling scenario, we can rewrite them in the equivalent form:

\begin{align}\label{gyni_new_form_nf}
\begin{aligned}
     \tilde{\mathcal{I}}_\text{GYNI} &= p(01|00) + p(10|00) + p(11|00) \\ &+ p(00|01) + p(01|01) + p(11|01) \\ & + p(00|10) + p(10|10) + p(11|10) \\ & + p(00|11) + p(01|11) + p(10|11) \geq 2 ,
\end{aligned}
\end{align}

\begin{align}\label{lgyni_new_form}
\begin{aligned}
     \tilde{\mathcal{I}}_\text{LGYNI} &=  p(01|01) +  p(11|01)  +  p(10|10) \\ &+  p(11|10) +  p(00|11) +  p(01|11) \\ &+  p(10|11)  \geq  1.
\end{aligned}
\end{align}

To see that $\tilde{\mathcal{I}}_\text{(L)GYNI}$ and $\mathcal{I}_\text{(L)GYNI}$ are indeed equivalent, one can use the normalization conditions $\sum_{a,b} p(a,b|x,y) = 1, \; \forall x,y$, to rewrite each term in inequalities Eq.(\ref{gyni}) and Eq.(\ref{lgyni}). Analysing the list of inequalities of the output-siganlling polytope, one can check that four versions of $\tilde{\mathcal{I}}_\text{LGYNI}$ equivalent under relabeling of inputs or parties are present, whereas no inequality equivalent to $\tilde{\mathcal{I}}_\text{GYNI}$ is present in the new polytope.

In Ref.~\cite{branciard2015simplest}, the maximal violation achieved for this inequality corresponds to\footnote{Note that due to the different form $\tilde{\mathcal{I}}_\text{LGYNI}$ of the LGYNI inequality in Eq.~(\ref{lgyni_new_form}), as compared to the one used in Ref.~\cite{branciard2015simplest}, it defines a causal lower bound, such that values smaller than the bound indicate violations. In conventions of Ref.~\cite{branciard2015simplest} the same maximal violation appears as $0.8194> 3/4$.} $0.7224 < 1$. Under our output-signalling framework, the optimal violation obtained via the see-saw optimization algorithm is $0.7780$. This reduced degree of violation is a direct consequence of the structural restrictions imposed on local operations. By constraining local strategies to measure-and-reprepare maps such as in Eq.~(\ref{measure_reprep_cj}), the corresponding local channels become entanglement-breaking, completely destroying any quantum coherence between a laboratory's input and output ports. Unlike general quantum instruments, which allow arbitrary coherent transformations and internal quantum memories, measure-and-reprepare operations restrict inter-party transmission strictly to classical measurement outcomes and state preparations. Because this constraint strictly reduces the space of achievable local instruments, a diminished capacity to generate noncausal correlations is naturally expected.

A paradigmatic example of a process matrix capable of generating noncausal correlations within the input-signalling framework of Ref.~\cite{branciard2015simplest} is given by (omitting superscript labels denoting input and output Hilbert spaces)
\begin{align}\label{eq: proc mat araujp}
    W = \frac{1}{4}\left[\mathbb{1}^{\otimes 4} + \frac{1}{\sqrt{2}}(ZZZ\mathbb{1} + Z\mathbb{1}XX)\right],
\end{align}
which yields an $\tilde{\mathcal{I}}_\text{LGYNI}$ violation of 0.8661 when Alice and Bob implement identical local instruments specified by
\begin{align}
    M^{X_I X_O}_{0\vert 0} &= 0, \\
    M^{X_I X_O}_{1\vert 0} &= 2\ket{\Phi^{+}} \bra{\Phi^{+}}, \\
    M^{X_I X_O}_{0\vert 1} &= \ket{0} \bra{0} \otimes \ket{0} \bra{0}, \\
    M^{X_I X_O}_{1\vert 1} &= \ket{1} \bra{1} \otimes \ket{0} \bra{0}.
\end{align}
with $X=A,B$. While the instruments associated to inputs $x,y=1$ do correspond to measure-and-reprepare channels, in order to reveal the noncausal correlations of this inseparable process matrix it's necessary to use non-entanglement breaking channels. In particular, in the strategy these channels correspond to not doing anything, that is, the same system entering Alice's or Bob's labs comes out unchanged when the inputs are $x,y=0$.

Within our measure-and-reprepare framework, we can reveal the noncausality of this process matrix making use of the output-signalling inequality
\begin{align}
\begin{aligned}
\mathcal{I}_4 &= p(10|00)  +  p(11|00) +2 p(00|01)  +  p(01|01) \\ &+2 p(10|01)  +  p(00|10)     +  p(01|10)  +  p(10|10)  \\ &+  p(01|11)  +  p(10|11)  +  p(11|11) \geq 2
 \end{aligned}
\end{align}
and the following states and measurements for each part. Alice applies the POVM's
\begin{align}
E^{A_I}_{0|0}
=
E^{A_I}_{1|1}
= \ket{1}\bra{1}, \quad
E^{A_I}_{1|0}
=
E^{A_I}_{0|1}
= \ket{0}\bra{0},
\end{align}
and reprepares the states according to the output of the measurement as
\begin{align}
\rho^{A_O}_0 = \ket{0}\bra{0}
, \quad
\rho^{A_O}_1 
= \ket{1}\bra{1}.
\end{align}
In turn, Bob implements the POVM's
\begin{align}
E^{B_I}_{0|0}
&= \frac{1}{2}\left(
\mathbb{1}-\frac{3X+Z}{\sqrt{10}}
\right),
&
E^{B_I}_{1|0}
&= \frac{1}{2}\left(
\mathbb{1}+\frac{3X+Z}{\sqrt{10}}
\right),\\
E^{B_I}_{0|1}
&= \frac{1}{2}\left(
\mathbb{1}+\frac{X+2Z}{\sqrt{5}}
\right),
&
E^{B_I}_{1|1}
&= \frac{1}{2}\left(
\mathbb{1}-\frac{X+2Z}{\sqrt{5}}
\right),
\end{align}
and signals his output through the states
\begin{align}
\rho^{B_O}_0 = \ket{+}\bra{+}, \quad 
\rho^{B_O}_1 = \ket{-}\bra{-}
\end{align}
The instruments of each party are built according to the measure-and-reprepare rule given in Eq. (\ref{measure_reprep_cj}). Then using the generalized Born's rule, the value of $\mathcal{I}_4$ (up to numerical precision) reads
\begin{align}
\mathcal{I}_4 = \frac{1}{4}(13 - \sqrt{10}-\sqrt{5}) \approx 1.9004\leq 2.
\end{align}

In the next section, we present one of the central findings of this work. Although the output-signalling framework yields smaller numerical violations for certain canonical process matrices (such as Eq.~(\ref{eq: proc mat araujp})), it exhibits a remarkable diagnostic advantage: it can uncover and certify the noncausality of process matrices whose indefinite causal order eludes detection by conventional input-signalling inequalities such as GYNI or LGYNI.

\subsection{Output signalling noncausal statistics unsimulable by input signalling strategies}
\label{sec: process matrix tailored for output signalling}

The analysis of causal inequalities established in this work shows that they define facets of a tighter polytope than that of the input-signalling scenario. Consequently, certain correlations classified as causal under input-signalling strategies appear as noncausal when evaluated within the output-signalling framework.

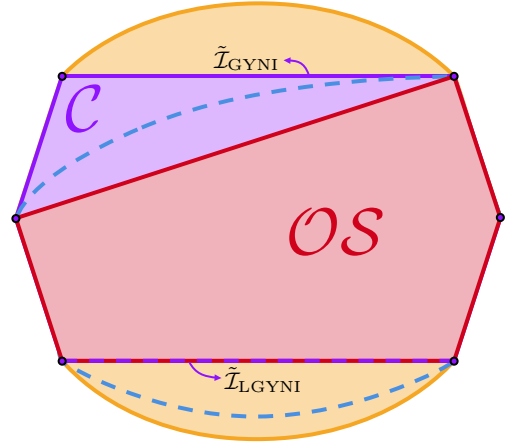
\begin{figure}
    \centering
    \input{images/boxes}
    \caption{Geometry of correlation sets with communication. The yellow convex set represents all valid achievable correlations with process matrices and arbitrary instruments, being a superset of the causal polytope without communication restrictions (denoted by $\mathcal{C}$). The latter is obtained through convex mixtures of deterministic strategies (purple dots) corresponding to the pair of Eqs. (\ref{inp_sing_expan}). The nontrivial facets of $\mathcal{C}$ are the families of inequalities $\tilde{\mathcal{I}}_\text{GYNI}$ and $\tilde{\mathcal{I}}_\text{LGYNI}$ (Eqs. (\ref{gyni_new_form_nf}) and (\ref{lgyni_new_form}), respectively). The output-signalling polytope (denoted by $\mathcal{OS}$) is represented by the red convex region, generated by the subset of the vertices of $\mathcal{C}$ corresponding to the output-signalling strategies, defined in Eqs. (\ref{out_sing_expan_a}) and (\ref{out_sing_expan_b}). Despite being generated by a subset of $\mathcal{C}$ vertices, the $\mathcal{OS}$ polytope has more facets in comparison with $\mathcal{C}$, a feature not appearing in Fig. \ref{fig:polytopes} due to its simplified two-dimensional description. The correlations obtainable with process matrix and measure-and-reprepare instruments (Eq. (\ref{measure_reprep_cj})) yield the convex set whose boundary is represented by the dashed blue line. As a consequence of this construction, there are correlations which are classified as causal for input-signalling strategies, but which yield noncausal statistics for output-signalling strategies: This is the the set of points contained in $\mathcal{C}$, which are inside of the blue dashed line and outside of $\mathcal{OS}$. The inequality $\tilde{\mathcal{I}}_\text{LGYNI}$  characterizes a facet of both $\mathcal{C}$ and $\mathcal{OS}$ polytopes, however $\tilde{\mathcal{I}}_\text{GYNI}$ is not a facet of $\mathcal{OS}$. Figure inspired by \cite{bavaresco2024indefinite}.  }
    \label{fig:polytopes}
\end{figure}

In this section, we provide explicit examples of such correlations for a family of process matrices. To certify these statistics, we augment the see-saw optimization procedure detailed in Sec.~\ref{sec: see saw}. During the minimization of a target inequality $\mathcal{I}_i$, in addition to the standard operational constraints, we strictly enforce compliance with \textit{all} 32 causal inequalities from the input-signalling scenario—namely, all valid relabelings of the GYNI and LGYNI inequalities \cite{branciard2015simplest}:
\begin{subequations}
\begin{align} 
    \min_{\{\mathcal{E}_\mathcal{A,B}, \mathcal{D_{A,B},} W\}} \quad &\mathcal{I}_i(\mathcal{E_{A,B}}, \mathcal{D_{A,B}}, W) \label{eq: input signalling ineq constraints1} \\
    \text{s.t.}\hspace{0.7cm} \quad &\mathcal{I}_\text{(L)GYNI} \geq \beta_\mathcal{C}. \label{eq: input signalling ineq constraints2}
\end{align}
\end{subequations}

This constrained optimization successfully yields multiple instances of correlations that are noncausal exclusively under output-signalling strategies, while remaining entirely causal from an input-signalling perspective. The majority of the optimal process matrices found in this manner are too algebraically cumbersome to present in their expanded form (Eq.~(\ref{eq: W basis decomposition})) and have therefore been made available in our online repository \cite{OutputSignallingData}. Nevertheless, by fixing a suitably chosen process matrix, we can specify a concrete set of states and POVMs that explicitly generate these strictly output-signalling noncausal statistics.

As an example, consider the following process matrix representing a superposition of distributed states and channels with memory:
\begin{align}\label{eq: goat process matrix}
\begin{aligned}
    W = \frac{1}{4} \left[ \mathbb{1}^{\otimes 4} + \frac{1}{\sqrt{2}}\left( YXY\mathbb{1} +  Z\mathbb{1}YX \right) \right].
\end{aligned}
\end{align} 
If Alice sends her outputs encoded in the states
\begin{align}
\rho^{A_O}_0 = \ket{+}\bra{+}, \quad  
\rho^{A_O}_1 = \ket{-}\bra{-},
\end{align}
being each of the related to the outcomes $a=0,1$ of the POVM's
\begin{subequations}
\begin{align}
E^{A_I}_{0|0} =
\ket{0}\bra{0}&, \quad E^{A_I}_{1|0} = \ket{1}\bra{1}, \\
E^{A_I}_{0|1} = \ket{+i}\bra{+i}&,
\quad E^{A_I}_{1|1} = 
\ket{-i}\bra{-i},
\end{align}
\end{subequations}
while Bob measures
\begin{subequations}
\begin{align}
E^{B_I}_{0|0} = 0&, \ E^{B_I}_{1|0} = \mathbb{1}, \\ E^{B_I}_{0|1} = 
\ket{-i}\bra{-i}&, \
E^{B_I}_{1|1} = \ket{+i}\bra{+i},
\end{align}
\end{subequations}
and reprepares the states
\begin{align}
\rho^{B_O}_0 = \ket{+}\bra{+},
\quad \rho^{B_O}_1 = \ket{-}\bra{-},
\end{align}
we can obtain a violation of the output-signalling facet-defining inequality $\mathcal{I}_{38}$:
\begin{align}\label{eq:ineq38}
\begin{aligned}
&\mathcal{I}_{38} = p(10|00)  +  p(11|00)  +  p(00|01)  +  p(01|01)  \\ &+p(00|10)  +  p(10|10)  +  p(00|11)  +  p(10|11) \geq 1. 
 \end{aligned}
\end{align}
Specifically, the violation is given by
\begin{align}
    \mathcal{I}_{38} = \frac{3-\sqrt{2}}{2} \approx 0.7929 \leq 1.
\end{align}

We emphasize that the above value is achieved under the constraint of input-signalling polytope inequalities (Equations (\ref{eq: input signalling ineq constraints1}) - (\ref{eq: input signalling ineq constraints2}) and their relabellings) being respected.

\begin{figure}
    \centering
    \includegraphics[width=1\linewidth]{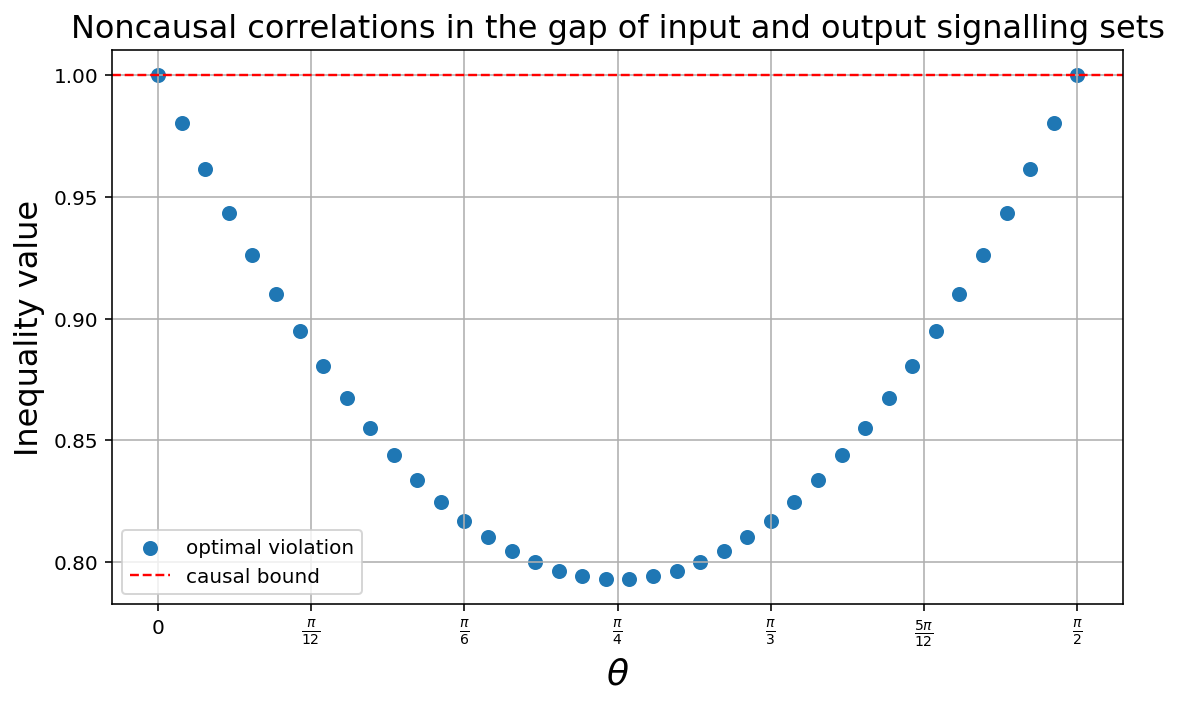}
    \caption{Statistics obtained for the violation of inequality~(\ref{eq:ineq38}) by the one-parameter family of process matrices defined by Eq. (\ref{eq: one parameter W}), while optimizing measure-and-reprepare instruments under the additional constraints of Eq. (\ref{eq: input signalling ineq constraints2}). Under this set of additional constraints, this represents a set of behaviours lying in the gap between the input and output signalling polytopes (see Fig. \ref{fig:polytopes}). For each value of $\theta$ we repeat the see-saw 200 times and select the smallest value. The data used to generate this plot can be found in \cite{OutputSignallingData}.
    }
   \label{fig: theta parametrization}
\end{figure}

Moreover, the process matrix given in Eq. (\ref{eq: goat process matrix}) can be extended to the one-parameter family of causal processes\footnote{The validity of $W(\theta)$ for all values of $\theta$ can be seen in a relatively simple way due to its Pauli decomposition. Both $YXY\mathbb{1}$ and $Z\mathbb{1}YX$ have support compatible with the linear process-matrix constraints and are traceless, so that $\text{Tr}[W(\theta)]=4 \ \forall \ \theta $. Moreover, these two Pauli strings anticommute, and when squared result in identity. Therefore
\begin{align}
[\cos(\theta)YXY\mathbb{1}+\sin(\theta)Z\mathbb{1}YX]^2=\mathbb{1}^{\otimes 4} \ \forall\ \theta.
\end{align}
Since any matrix that squares to the identity has eigenvalues $\pm1$ it follows that the eigenvalues of $W(\theta)$ are $0$ and $\nicefrac{1}{2}$, that is $W(\theta)\geq0$ in this interval.}
\begin{align}\label{eq: one parameter W}
    W(\theta)= \frac{1}{4} \left[ \mathbb{1}^{\otimes 4} +  \cos(\theta)YXY\mathbb{1} + \sin(\theta) Z\mathbb{1}YX  \right],
\end{align}
which are causally separable for $\theta=0,\nicefrac{\pi}{2}$ and nonseparable inside the interval $\theta \in (0,\nicefrac{\pi}{2})$.  Figure \ref{fig: theta parametrization} illustrates this example of noncausal correlations achievable with output signalling and measure-and-reprepare strategies, not captured by the (L)GYNI inequalities.

\subsection{State misalignment thresholds allowing noncausal statistics}
\label{sec: orthogonality thresholds}

In measure-and-reprepare instruments, encoding a classical bit $a$ into an output state $\rho_a \in \mathcal{L}(\mathcal{H}^{A_O})$ is susceptible to experimental imperfections at both ends of the communication protocol. Crucially, states encoding distinct bits must be as distinguishable as possible. If a pair of states $(\rho_a, \rho_{a^{\prime}})$ used to encode distinct bits $(a,a^{\prime})$ possesses a non-zero overlap, the receiving party has a proportional probability of misidentifying the transmitted signal. Consequently, as observed in our previous optimizations, maximal inequality violations typically require each party to prepare perfectly orthogonal states.

To illustrate how such imperfections can be addressed within our framework, we employ the numerical see-saw method described in Sec. \ref{sec: see saw} to evaluate the impact of limited state distinguishability. We model this by assuming Alice encodes her bits using a pair of pure states characterized by a misalignment angle $\phi$:
\begin{align}\label{eq: imperfect encoding1}
    \rho_0^{A_O} &= \ket{0}\bra{0}, \quad \rho_1^{A_O}(\phi) = \ket{\psi^{A}(\phi)}\bra{\psi^{A}(\phi)},
\end{align}
with the misaligned state $\ket{\psi^{A}(\phi)}$ being
\begin{align}
    \ket{\psi^{A}(\phi)} = \cos\left(\frac{\phi}{2}\right)\ket{0} + \sin\left(\frac{\phi}{2}\right) \ket{1}.
\end{align}
We also assume that Bob outputs misaligned states, which in an ideal configuration would correspond to the $\{\ket{+},\ket{-}\}$ basis:
\begin{align}
    \rho_0^{B_O} &= \ket{+}\bra{+}, \quad \rho_1^{B_O}(\phi) = \ket{\psi^{B}(\phi)}\bra{\psi^{B}(\phi)}.
\end{align}
For simplicity, we assume his imperfect states are misaligned by the same angle $\phi$:
\begin{align}\label{eq: imperfect encoding 2}
    \ket{\psi^{B}(\phi)}=\cos\left(\frac{\phi}{2}\right)\ket{+} + \sin\left(\frac{\phi}{2}\right) \ket{-}.
\end{align}

If $\phi=0$, Alice and Bob encode their classical bits into identical states; therefore, the output signals carry no information, precluding the detection of any noncausal statistics. However, as the state overlap decreases, the encoding fidelity progressively improves, allowing more noncausality to be witnessed. The optimal scenario is reached at $\phi = \pi$, where the signals are encoded into perfectly orthogonal states, ensuring no information is lost during transmission.

\begin{figure}
    \centering
    \includegraphics[width=1\linewidth]{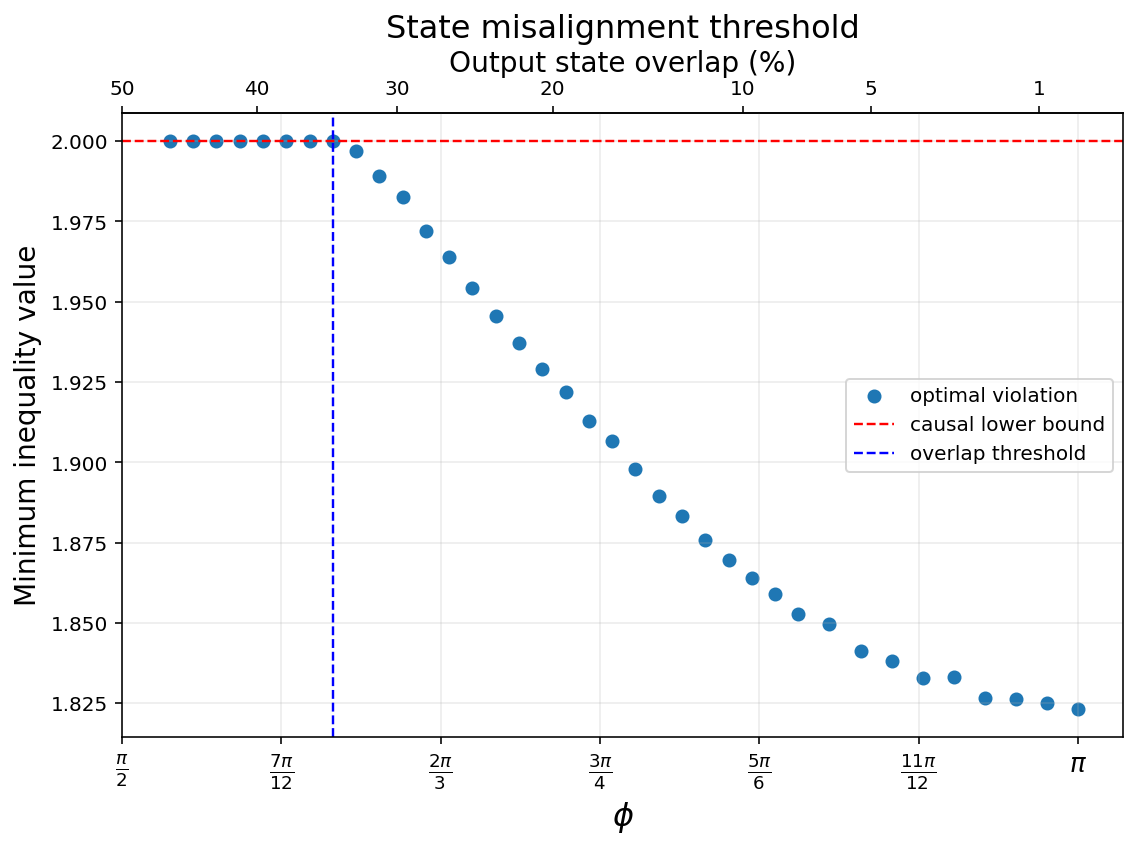}
    \caption{Plot showcasing the state-misalignment threshold, parametrized by the angle $\phi$, for reprepared states to reveal noncausal correlations through inequality $\mathcal{I}_{11}$. The red dashed line corresponds to the causal bound of the inequality, while the blue dashed line indicates the threshold above which a violation is obtained. For each value of $\phi$, the see-saw optimization was repeated 100 times, and the data points correspond to the lowest values found. The upper horizontal axis shows the percentage overlap between the states used to encode distinct outputs, given by $\cos^2(\phi/2)$. In the case of inequality $\mathcal{I}_{11}$, a violation requires $\phi \gtrsim 1.92\,\mathrm{rad}$ ($\approx 110^\circ$), corresponding to an output-state overlap of approximately $32.89\%$.
    }
   \label{fig: state distinguishability}
\end{figure}

To analyze the behavior of noncausal statistics under this imperfect state encoding model given in Equations (\ref{eq: imperfect encoding1}) - (\ref{eq: imperfect encoding 2}), we evaluate the inequality
\begin{align}
\begin{aligned}
\mathcal{I}_{11} &=
 p(01|00)  +  p(10|00)  +  p(11|00)  +  p(00|01) \\ & +  p(10|01)    +  p(00|10)  +  p(01|10)  +  p(10|10) \\ & +  p(01|11) +2 p(10|11) +2 p(11|11) \geq 2,
 \end{aligned}
\end{align}
together with the see-saw optimization detailed in Sec. \ref{sec: see saw}. By fixing the states for varying values of the misalignment angle $\phi$, and optimizing over the process matrices and POVMs, our framework identifies a state-overlap threshold required to extract noncausal statistics. For inequality $\mathcal{I}_{11}$, the maximum tolerable state overlap was found to be approximately $32.89\%$. The minimal inequality value achieved as a function of the misalignment parameter $\phi$ is illustrated in Fig. \ref{fig: state distinguishability}.

\subsection{General violations of output-signalling inequalities}
\label{sec: violation bounds}

Here we exhibit the violations that could be found for the inequalities characterizing the output signalling causal polytope throught the see-saw technique explained in Sec. \ref{sec: see saw}. Due to the large number of inequalities, for convenience we display their explicit expression in Appendix , while the matrix representation for the process matrix and instruments used to obtain their violations can be found in the repository \cite{OutputSignallingData}.

Interestingly, the see-saw technique detailed in Sec. \ref{sec: see saw} finds no violation for some of the output-signalling inequalities. This absence of violation points to two possibilities. First, the inequalities might be facets of both the output-signalling polytope and the set of process matrix correlations. In this scenario, the maximum value attainable via output-signalling strategies perfectly coincides with the bound provided by the process matrix framework (Sec. \ref{sec: process matrix formalism}). Second, it is possible that a violation exists, but the see-saw algorithm fails to reach it.

\begin{table}
\begin{ruledtabular}
\begin{tabular}{ccc}
\makecell{Inequality \\ (Appendix \ref{app: inequality list})} & Causal lower bound & \makecell{See-saw value \\ (Sec. \ref{sec: see saw})}\\
\hline
$\mathcal{I}_{1}$ & 2 & 1.8589 \\
$\mathcal{I}_{2}$ & 2 & 1.8226 \\
$\mathcal{I}_{3}$ & 2 & 1.8589 \\
$\mathcal{I}_{4}$ & 2 & 1.8226 \\
$\mathcal{I}_{5}$ & 2 & 1.8150 \\
$\mathcal{I}_{6}$ & 2 & 1.7786 \\
$\mathcal{I}_{7}$ & 2 & 1.7786 \\
$\mathcal{I}_{8}$ & 2 & 1.8226 \\
$\mathcal{I}_{9}$ & 2 & 1.8151 \\
$\mathcal{I}_{10}$ & 2 & 1.7786 \\
$\mathcal{I}_{11}$ & 2 & 1.8226 \\
$\mathcal{I}_{12}$ & 2 & 1.8969 \\
$\mathcal{I}_{13}$ & 2 & 1.8589 \\
$\mathcal{I}_{14}$ & 2 & 1.8150 \\
$\mathcal{I}_{15}$ & 2 & 1.8590 \\
$\mathcal{I}_{16}$ & 2 & 1.8227 \\
$\mathcal{I}_{17}$ & 2 & 1.8969 \\
$\mathcal{I}_{18}$ & 1 & -  \\
$\mathcal{I}_{19}$ & 1 & - \\
$\mathcal{I}_{20} \left( \tilde{\mathcal{I}}_\text{LGYNI} \right)$ & 1 & 0.7780 \\
$\mathcal{I}_{21}$ & 1 & 0.7500 \\
$\mathcal{I}_{22}$ & 1 & 0.7780 \\
$\mathcal{I}_{23}$ & 1 & 0.7500 \\
$\mathcal{I}_{24}$ & 1 & -  \\
$\mathcal{I}_{25}$ & 1 & 0.7500 \\
$\mathcal{I}_{26}$ & 1 & 0.7782 \\
$\mathcal{I}_{27}$ & 1 & 0.7500 \\
$\mathcal{I}_{28}$ & 1 & - \\
$\mathcal{I}_{29}$ & 1 & - \\
$\mathcal{I}_{30}$ & 1 & 0.9345 \\
$\mathcal{I}_{31}$ & 1 & 0.8722 \\
$\mathcal{I}_{32}$ & 1 & 0.8722 \\
$\mathcal{I}_{33}$ & 1 & 0.9345 \\
$\mathcal{I}_{34}$ & 1 & 0.7500 \\
$\mathcal{I}_{35}$ & 1 & 0.8722 \\
$\mathcal{I}_{36}$ & 1 & 0.9345 \\
$\mathcal{I}_{37}$ & 1 & 0.8722 \\
$\mathcal{I}_{38}$ & 1 & 0.7500 \\
\end{tabular}
\end{ruledtabular}
\caption{Upper bounds for the violations of the causal inequalities of the ouput-signalling polytope. The matrix representation of the instruments and process matrices leading to such values can be found in the repository \cite{OutputSignallingData}.}
\end{table}

\section{Conclusion}
\label{sec: conclusion}

In this work, we introduced and analyzed an output-signalling framework for bipartite indefinite causal order, imposing a fundamental restriction on local operations: parties may only communicate the classical outcomes of their local instruments, strictly prohibiting the direct transmission of their input settings. By geometrically characterizing the causal polytope under this assumption, we identified $38$ distinct equivalence families of nontrivial causal inequalities. Crucially, we demonstrated that these classical bounds remain entirely valid for all quantum processes operating under a definite causal order; thus, their violation serves as a genuine, robust witness of indefinite causal order (ICO).

Employing a see-saw optimization technique based on semidefinite programming, we showed that parties restricted to measure-and-reprepare instruments, which act as entanglement-breaking channels and destroy input-output quantum coherence, can nevertheless generate strong noncausal correlations. A central result of our analysis is the discovery that the output-signalling causal polytope is strictly tighter than its input-signalling counterpart. Consequently, our framework is uniquely capable of certifying the noncausality of process matrices that appear entirely causal to conventional input-signalling witnesses, such as the GYNI and LGYNI inequalities \cite{branciard2015simplest}. Furthermore, by modeling the encoding of classical bits into non-orthogonal quantum states, we have shown that these noncausal statistics display a reasonable tolerance to measurement imperfections, withstanding up to a $\approx 33\%$ state overlap. 

These findings open several promising avenues for future research. While our present analysis was restricted to qubit Hilbert spaces and binary inputs/outputs, generalizing this framework to higher-dimensional systems (qudits) and larger input/output cardinalities is a natural next step. Based on analogous results in input-signalling scenarios \cite{branciard2015simplest}, we anticipate that higher-dimensional spaces will yield even more robust noncausal correlations. Another direction would be to understand whether the measure-and-reprepare instruments could be optimal among all possible entanglement-breaking channels. Finally, we leave as a compelling open question whether the output-signalling causal inequalities derived here can reveal noncausal statistics in purifiable process matrices \cite{araujo2017purification} (which would require at least a tripartite scenario \cite{yokojima2021consequences}), a critical theoretical step toward understanding the fundamental physical realizability of these exotic causal structures.

\section{Acknowledgements} 

We thank Gláucia Murta, Jessica Bavaresco, Marco Túlio Quintino and Veronika Baumann for fruitful discussions. We acknowledge funding from the Austrian Research Promotion Agency (FFG) through the Project NSPT-QKD FO999915265, the Simons Foundation (Grant No. 1023171, R.C.), the Brazilian National Council for Scientific and Technological Development (CNPq, Grants No. 403181/2024-0, 301687/2025-0, 150631/2026-0, R.C. and 309603/2026-9, 404274/2023-4, D.M.) the Financiadora de Estudos e Projetos (Grant No. 1699/24 IIF-FINEP). RC thanks the Technical University of Denmark for its hospitality, where part of this work was carried out during a guest professorship supported by the Otto M\o nsted Foundation. We also thank the High-Performance Computing Center (NPAD) at UFRN for providing computational resources.

\appendix

\section{Representatives of facet-defining inequalities of the output-signalling polytope \label{app: inequality list}}

Here we present the full list of representatives of each class of output-signalling causal inequalities characterizing the facets of the output-signalling polytope. The first 38 inequalities represent nontrivial classes, while the last three inequalities represent the classes of non-negative probabilities.

\begin{align}
\begin{aligned}
\mathcal{I}_1 &= p(10|00)  +  p(11|00)  +  p(00|01) +2 p(01|01) \\ &+2 p(11|01)  +  p(01|10)    +  p(10|10)  +  p(11|10)   \\  &+p(00|11)  +  p(01|11)  +  p(10|11) \geq 2 
 \end{aligned}
\end{align}

\begin{align}
\begin{aligned}
\mathcal{I}_2 &= p(10|00)  +  p(11|00)  +  p(00|01) +2 p(01|01) \\ &+2 p(11|01)  +  p(00|10)     +  p(01|10)  +  p(11|10)   \\ &+p(00|11)  +  p(10|11)  +  p(11|11) \geq 2 
 \end{aligned}
\end{align}

\begin{align}
\begin{aligned}
\mathcal{I}_3 &= p(10|00)  +  p(11|00) +2 p(00|01)  +  p(01|01) \\ &+2 p(10|01)  +  p(00|10)     +  p(10|10)  +  p(11|10) \\ &+  p(00|11)  +  p(01|11)  +  p(11|11) \geq 2
 \end{aligned}
\end{align}

\begin{align}
\begin{aligned}
\mathcal{I}_4 &= p(10|00)  +  p(11|00) +2 p(00|01)  +  p(01|01) \\ &+2 p(10|01)  +  p(00|10)     +  p(01|10)  +  p(10|10)  \\ &+  p(01|11)  +  p(10|11)  +  p(11|11) \geq 2
 \end{aligned}
\end{align}

\begin{align}
\begin{aligned}
\mathcal{I}_5 &= p(10|00)  +  p(11|00) +2 p(00|01) +2 p(01|01) \\ &+  p(11|01)  +  p(00|10)   +  p(01|10)  +  p(10|10) \\ &+2 p(00|11) +2 p(10|11) \geq 2
 \end{aligned}
\end{align}

\begin{align}
\begin{aligned}
\mathcal{I}_6 &=  p(10|00)  +  p(11|00) +2 p(00|01) +2 p(01|01) \\ 
&+2 p(11|01)  +  p(00|10) +  p(01|10)  +  p(11|10) \\ 
&+  p(00|11)  +  p(10|11) \geq 2
 \end{aligned}
\end{align}

\begin{align}
\begin{aligned}
\mathcal{I}_7 &=
 p(10|00)  +  p(11|00) +2 p(00|01) +2 p(01|01) \\ &+2 p(10|01)  +  p(00|10) +  p(01|10) +p(10|10)  \\ &+  p(01|11)  +  p(11|11) \geq 2 
 \end{aligned}
\end{align}

\begin{align}
\begin{aligned} 
\mathcal{I}_8 &=
 2 p(10|00) +2 p(11|00)  +  p(00|01) \\  &+2 p(11|01)  +  p(01|10)  +2 p(01|01)   +  p(10|10)  \\ &+  p(11|10) +  p(00|11)  +  p(10|11) \geq 2 
 \end{aligned}
\end{align}

\begin{align}
\begin{aligned}
\mathcal{I}_{9} &=
 2 p(10|00) +2 p(11|00) +2 p(00|01) \\& +  p(01|01) +2 p(10|01)  +  p(00|10)    +  p(10|10) \\ &+  p(11|10)  +  p(01|11)  +  p(11|11) \geq 2 
 \end{aligned}
\end{align}

\begin{align}
\begin{aligned}
\mathcal{I}_{10} &= p(01|00)  +  p(10|00)  +  p(11|00)\\ &  +  p(00|01)  +  p(10|01)  +  p(00|10) +  p(01|10)  \\ &+2 p(01|11) +2 p(10|11)  +2 p(11|11) \geq 2 
 \end{aligned}
\end{align}

\begin{align}
\begin{aligned}
\mathcal{I}_{11} &=
 p(01|00)  +  p(10|00)  +  p(11|00)  +  p(00|01) \\ & +  p(10|01)    +  p(00|10)  +  p(01|10)  +  p(10|10) \\ & +  p(01|11) +2 p(10|11) +2 p(11|11) \geq 2 
 \end{aligned}
\end{align}

\begin{align}
\begin{aligned}
&\mathcal{I}_{12} =
 p(01|00)  +  p(10|00)  +  p(11|00)  +  p(00|01) \\ & +  p(01|01)    +  p(10|01)  +  p(00|10)  +  p(01|10) \\ &  +  p(10|10)  +  p(01|11)  +  p(10|11)  +  p(11|11) \geq 2 
 \end{aligned}
\end{align}

\begin{align}
\begin{aligned}
&\mathcal{I}_{13} =
 p(01|00)  +  p(10|00)  +  p(11|00)  +  p(00|01) \\ & +  p(01|01)    +  p(10|01) +2 p(00|10)  +2 p(10|10) \\ & +   p(11|10)  +  p(00|11)  +  p(01|11) \geq 2 
 \end{aligned}
\end{align}

\begin{align}
\begin{aligned}
&\mathcal{I}_{14} =
 p(01|00) +2 p(10|00) +2 p(11|00)   \\ & +  p(00|01) +  p(01|01) +2 p(00|10)   +2 p(10|10) \\ & +  p(00|11)   +  p(10|11)  +  p(11|11) \geq 2 
 \end{aligned}
\end{align}

\begin{align}
\begin{aligned}
&\mathcal{I}_{15} =
 p(01|00) +2 p(10|00) +2 p(11|00)   \\ & +  p(01|01)  +  p(11|01)    +  p(00|10)  +  p(10|10) \\ & +  p(00|11)  +  p(10|11)  +  p(11|11) +  p(00|01) \geq 2 
 \end{aligned}
\end{align}

\begin{align}
\begin{aligned}
&\mathcal{I}_{16} =
 p(00|00)  +  p(10|00)  +  p(00|01) +  p(01|01)  \\ &  +  p(11|01) +2 p(00|10)   +2 p(01|10)  +  p(11|10) \\ &  +  p(00|11)  +  p(10|11)  +  p(11|11) \geq 2 
 \end{aligned}
\end{align}

\begin{align}
\begin{aligned}
&\mathcal{I}_{17} =
 p(00|00)  +  p(10|00)  +  p(11|00)  +  p(00|01) \\ & +  p(01|01)  +  p(11|01)   +  p(00|10)  +  p(01|10) \\ & +  p(11|10)  +  p(00|11)  +  p(10|11)  +  p(11|11) \geq 2 
 \end{aligned}
\end{align}

\begin{align}
\begin{aligned}
\mathcal{I}_{18} &=
 p(01|10)  +  p(10|10)  +  p(11|10) \\ &  +  p(00|11)  +  p(01|11)  +  p(11|11) \geq 1 
 \end{aligned}
\end{align}

\begin{align}
\begin{aligned}
\mathcal{I}_{19} &=
 p(00|10)  +  p(10|10)  +  p(11|10) \\ &  +  p(00|11)  +  p(01|11)  +  p(10|11) \geq 1 
 \end{aligned}
\end{align}

\begin{align}
\begin{aligned}
\mathcal{I}_{20} &=
 p(01|01)  +  p(11|01)  +  p(10|10) +  p(11|10)  \\ &   +  p(00|11)  +  p(01|11)  +  p(10|11) \geq 1 
 \end{aligned}
\end{align}

\begin{align}
\begin{aligned}
\mathcal{I}_{21} &=
 p(01|01)  +  p(11|01)  +  p(01|10) + p(10|10) \\ &   +  p(11|10)  +  p(00|11)  +  p(01|11) \geq 1 
 \end{aligned}
\end{align}

\begin{align}
\begin{aligned}
\mathcal{I}_{22} &=
 p(01|01)  +  p(11|01)  +  p(00|10) +  p(01|10) \\ &   +  p(00|11)  +  p(10|11)  +  p(11|11) \geq 1 
 \end{aligned}
\end{align}

\begin{align}
\begin{aligned}
\mathcal{I}_{23} &=
 p(01|01)  +  p(11|01)  +  p(00|10) +  p(01|10) \\ &   +  p(11|10)  +  p(10|11)  +  p(11|11) \geq 1 
 \end{aligned}
\end{align}

\begin{align}
\begin{aligned}
\mathcal{I}_{24} &=
 p(01|01)  +  p(10|01)  +  p(11|01) +  p(01|10)  \\ &  +  p(10|10)  +  p(11|10)  +  p(00|11) \geq 1 
 \end{aligned}
\end{align}

\begin{align}
\begin{aligned}
\mathcal{I}_{25} &=
 p(00|01)  +  p(10|01)  +  p(00|10) +  p(10|10)  \\ &  +  p(11|10)  +  p(00|11)  +  p(01|11) \geq 1 
 \end{aligned}
\end{align}

\begin{align}
\begin{aligned}
\mathcal{I}_{26} &=
 p(00|01)  +  p(10|01)  +  p(00|10) +  p(01|10) \\ &   +  p(01|11)  +  p(10|11)  +  p(11|11) \geq 1 
 \end{aligned}
\end{align}

\begin{align}
\begin{aligned}
\mathcal{I}_{27} &=
 p(00|01)  +  p(10|01)  +  p(00|10) +  p(01|10) \\ &   +  p(10|10)  +  p(10|11)  +  p(11|11) \geq 1 
 \end{aligned}
\end{align}

\begin{align}
\begin{aligned}
\mathcal{I}_{28} &=
 p(00|01)  +  p(10|01)  +  p(11|01)  +  p(00|10) \\ &  +  p(10|10)  +  p(11|10)  +  p(01|11) \geq 1 
 \end{aligned}
\end{align}

\begin{align}
\begin{aligned}
\mathcal{I}_{29} &=
 p(00|01)  +  p(01|01)  +  p(10|01)  +  p(00|10)  \\ & +  p(01|10)  +  p(10|10)  +  p(11|11) \geq 1 
 \end{aligned}
\end{align}

\begin{align}
\begin{aligned}
&\mathcal{I}_{30} =
 p(11|00)  +  p(00|01)  +  p(01|01)  +  p(00|10) \\ & +  p(10|10)  +  p(00|11)  +  p(01|11)  +  p(10|11) \geq 1 
 \end{aligned}
\end{align}

\begin{align}
\begin{aligned}
&\mathcal{I}_{31} =
 p(11|00)  +  p(00|01)  +  p(01|01)  +  p(00|10) \\ & +  p(01|10)  +  p(10|10)  +  p(00|11)  +  p(10|11) \geq 1 
 \end{aligned}
\end{align}

\begin{align}
\begin{aligned}
&\mathcal{I}_{32} =
 p(10|00)  +  p(11|00)  +  p(01|01)  +  p(00|10) \\ & +  p(10|10)  +  p(00|11)  +  p(10|11)  +  p(11|11) \geq 1 
 \end{aligned}
\end{align}

\begin{align}
\begin{aligned}
&\mathcal{I}_{33} =
 p(10|00)  +  p(11|00)  +  p(01|01)  +  p(00|10) \\ & +  p(10|10)  +  p(11|10)  +  p(00|11)  +  p(10|11) \geq 1 
 \end{aligned}
\end{align}

\begin{align}
\begin{aligned}
&\mathcal{I}_{34} =
 p(10|00)  +  p(11|00)  +  p(01|01)  +  p(11|01) \\ & +  p(10|10)  +  p(11|10)  +  p(00|11)  +  p(10|11) \geq 1 
 \end{aligned}
\end{align}

\begin{align}
\begin{aligned}
&\mathcal{I}_{35} =
 p(10|00)  +  p(11|00)  +  p(01|01)  +  p(11|01) \\ & +  p(01|10)  +  p(10|10)  +  p(11|10)  +  p(00|11) \geq 1 
 \end{aligned}
\end{align}

\begin{align}
\begin{aligned}
&\mathcal{I}_{36} =
 p(10|00)  +  p(11|00)  +  p(00|01)  +  p(01|10) \\ & +  p(10|10)  +  p(11|10)  +  p(01|11)  +  p(11|11) \geq 1 
 \end{aligned}
\end{align}

\begin{align}
\begin{aligned}
&\mathcal{I}_{37} =
 p(10|00)  +  p(11|00)  +  p(00|01)  +  p(10|01) \\ & +  p(00|10)  +  p(10|10)  +  p(11|10)  +  p(01|11) \geq 1 
 \end{aligned}
\end{align}

\begin{align}
\begin{aligned}
&\mathcal{I}_{38} = p(10|00)  +  p(11|00)  +  p(00|01)  +  p(01|01)  \\ &+p(00|10)  +  p(10|10)  +  p(00|11)  +  p(10|11) \geq 1 
 \end{aligned}
\end{align}

\begin{align}
\begin{aligned}
\mathcal{I}_{39} &= p(11|11) \geq 0 
 \end{aligned}
\end{align}

\begin{align}
\begin{aligned}
\mathcal{I}_{40} &= p(10|11) \geq 0 
 \end{aligned}
\end{align}

\begin{align}
\begin{aligned}
\mathcal{I}_{41} &= p(00|11) \geq 0 
 \end{aligned}
\end{align}

\bibliography{main.bib}

\end{document}

%% file: images/ICO_scenario.tex
\tikzset{every picture/.style={line width=0.75pt}} 

\begin{tikzpicture}[x=0.75pt,y=0.75pt,yscale=-1,xscale=1,scale=0.5]

\draw   (583.63,161.28) .. controls (583.63,140.69) and (599.71,124) .. (619.56,124) .. controls (639.41,124) and (655.5,140.69) .. (655.5,161.28) .. controls (655.5,181.87) and (639.41,198.56) .. (619.56,198.56) .. controls (599.71,198.56) and (583.63,181.87) .. (583.63,161.28) -- cycle ;
\draw  [color={rgb, 255:red, 0; green, 0; blue, 0 }  ,draw opacity=1 ][fill={rgb, 255:red, 74; green, 144; blue, 226 }  ,fill opacity=1 ] (619.56,198.56) -- (634.08,190.21) -- (634.03,207) -- cycle ;
\draw  [fill={rgb, 255:red, 74; green, 144; blue, 226 }  ,fill opacity=1 ] (619.56,198.56) -- (605.03,206.88) -- (605.11,190.09) -- cycle ;
\draw  [fill={rgb, 255:red, 74; green, 144; blue, 226 }  ,fill opacity=1 ] (613.99,198.56) .. controls (613.99,195.37) and (616.49,192.78) .. (619.56,192.78) .. controls (622.64,192.78) and (625.13,195.37) .. (625.13,198.56) .. controls (625.13,201.76) and (622.64,204.34) .. (619.56,204.34) .. controls (616.49,204.34) and (613.99,201.76) .. (613.99,198.56) -- cycle ;
\draw[very thick, color={rgb, 255:red, 208; green, 2; blue, 27 } ]   (181.5,117) -- (269.5,117) -- (269.5,205) -- (181.5,205) -- cycle ;
\draw  [color={rgb, 255:red, 126; green, 211; blue, 33 }, draw opacity=1, fill opacity=1, black, ultra thick]  (182,12) -- (539.25,12) -- (539.25,52) -- (182,52) -- cycle ;
\draw  [color={rgb, 255:red, 126; green, 211; blue, 33 }, draw opacity=1, fill opacity=1, line width=1.5, black]  (181.5,271) -- (537.75,271) -- (537.75,311) -- (181.5,311) -- cycle ;
\draw  [color={rgb, 255:red, 126; green, 211; blue, 33 }, draw opacity=1, fill opacity=1, black, ultra thick]  (439.75,36) -- (439.75,282.67) -- (279.75,282.67) -- (279.75,36) -- cycle ;
\draw    (138.67,179.67) -- (170.42,179.67) ;
\draw [shift={(173.42,179.67)}, rotate = 180] [fill={rgb, 255:red, 0; green, 0; blue, 0 }  ][line width=0.08]  [draw opacity=0] (5.36,-2.57) -- (0,0) -- (5.36,2.57) -- cycle    ;
\draw    (139.5,141.5) -- (171.25,141.5) ;
\draw [shift={(136.5,141.5)}, rotate = 0] [fill={rgb, 255:red, 0; green, 0; blue, 0 }  ][line width=0.08]  [draw opacity=0] (5.36,-2.57) -- (0,0) -- (5.36,2.57) -- cycle    ;
\draw    (545.58,140.17) -- (577.33,140.17) ;
\draw [shift={(580.33,140.17)}, rotate = 180] [fill={rgb, 255:red, 0; green, 0; blue, 0 }  ][line width=0.08]  [draw opacity=0] (5.36,-2.57) -- (0,0) -- (5.36,2.57) -- cycle    ;
\draw    (547.75,179.67) -- (579.5,179.67) ;
\draw [shift={(544.75,179.67)}, rotate = 0] [fill={rgb, 255:red, 0; green, 0; blue, 0 }  ][line width=0.08]  [draw opacity=0] (5.36,-2.57) -- (0,0) -- (5.36,2.57) -- cycle    ;
\draw    (226.8,110.8) -- (226.8,61) ;
\draw [shift={(226.8,58)}, rotate = 90] [fill={rgb, 255:red, 0; green, 0; blue, 0 }  ][line width=0.08]  [draw opacity=0] (5.36,-2.57) -- (0,0) -- (5.36,2.57) -- cycle    ;
\draw    (226.8,264.8) -- (226.8,215) ;
\draw [shift={(226.8,212)}, rotate = 90] [fill={rgb, 255:red, 0; green, 0; blue, 0 }  ][line width=0.08]  [draw opacity=0] (5.36,-2.57) -- (0,0) -- (5.36,2.57) -- cycle    ;
\draw    (495.8,110.8) -- (495.8,61) ;
\draw [shift={(495.8,58)}, rotate = 90] [fill={rgb, 255:red, 0; green, 0; blue, 0 }  ][line width=0.08]  [draw opacity=0] (5.36,-2.57) -- (0,0) -- (5.36,2.57) -- cycle    ;
\draw    (495.8,264.8) -- (495.8,215) ;
\draw [shift={(495.8,212)}, rotate = 90] [fill={rgb, 255:red, 0; green, 0; blue, 0 }  ][line width=0.08]  [draw opacity=0] (5.36,-2.57) -- (0,0) -- (5.36,2.57) -- cycle    ;
\draw  [color={rgb, 255:red, 255; green, 255; blue, 255 }  ,draw opacity=1 ][fill={rgb, 255:red, 255; green, 255; blue, 255 }  ,fill opacity=1 ] (276.75,274) -- (443.75,274) -- (443.75,299.3) -- (277.5,299.3) -- cycle ;
\draw  [color={rgb, 255:red, 255; green, 255; blue, 255 }  ,draw opacity=1 ][fill={rgb, 255:red, 255; green, 255; blue, 255 }  ,fill opacity=1 ] (277,274.6) -- (444,274.6) -- (444,299.05) -- (277.75,299.05) -- cycle ;
\draw  [color={rgb, 255:red, 255; green, 255; blue, 255 }  ,draw opacity=1 ][fill={rgb, 255:red, 255; green, 255; blue, 255 }  ,fill opacity=1 ] (276.75,23.6) -- (440.75,23.6) -- (443.75,48.9) -- (277.5,48.9) -- cycle ;
\draw[very thick, color={rgb, 255:red, 74; green, 144; blue, 226 }]   (451.5,118) -- (539.5,118) -- (539.5,206) -- (451.5,206) -- cycle ;
\draw  [fill={rgb, 255:red, 255; green, 255; blue, 255 }  ,fill opacity=1 ] (41.25,161.41) .. controls (41.25,141.12) and (56.77,124.67) .. (75.91,124.67) .. controls (95.05,124.67) and (110.57,141.12) .. (110.57,161.41) .. controls (110.57,181.7) and (95.05,198.15) .. (75.91,198.15) .. controls (56.77,198.15) and (41.25,181.7) .. (41.25,161.41) -- cycle ;
\draw  [draw opacity=0] (41.24,161.41) .. controls (41.24,161.42) and (41.24,161.43) .. (41.24,161.43) .. controls (36.59,176.2) and (26.11,186.67) .. (14.82,189.41) -- (13.1,151.47) -- cycle ; \draw   (41.24,161.41) .. controls (41.24,161.42) and (41.24,161.43) .. (41.24,161.43) .. controls (36.59,176.2) and (26.11,186.67) .. (14.82,189.41) ;  
\draw  [draw opacity=0] (41.26,161.4) .. controls (41.26,161.4) and (41.26,161.41) .. (41.26,161.41) .. controls (37.2,169.2) and (28.99,173.28) .. (20.11,172.92) -- (17.56,147.53) -- cycle ; \draw   (41.26,161.4) .. controls (41.26,161.4) and (41.26,161.41) .. (41.26,161.41) .. controls (37.2,169.2) and (28.99,173.28) .. (20.11,172.92) ;  
\draw  [draw opacity=0] (41.25,161.39) .. controls (41.24,161.4) and (41.24,161.4) .. (41.23,161.41) .. controls (35.9,165.64) and (29.1,166.63) .. (22.53,164.77) -- (24.98,138.47) -- cycle ; \draw   (41.25,161.39) .. controls (41.24,161.4) and (41.24,161.4) .. (41.23,161.41) .. controls (35.9,165.64) and (29.1,166.63) .. (22.53,164.77) ;  
\draw  [draw opacity=0] (110.74,161.6) .. controls (110.74,161.61) and (110.75,161.61) .. (110.75,161.62) .. controls (115.4,176.38) and (125.88,186.85) .. (137.17,189.58) -- (138.9,151.65) -- cycle ; \draw   (110.74,161.6) .. controls (110.74,161.61) and (110.75,161.61) .. (110.75,161.62) .. controls (115.4,176.38) and (125.88,186.85) .. (137.17,189.58) ;  
\draw  [draw opacity=0] (110.75,161.57) .. controls (110.75,161.57) and (110.75,161.58) .. (110.76,161.59) .. controls (114.82,169.38) and (123.02,173.47) .. (131.89,173.12) -- (134.44,147.71) -- cycle ; \draw   (110.75,161.57) .. controls (110.75,161.57) and (110.75,161.58) .. (110.76,161.59) .. controls (114.82,169.38) and (123.02,173.47) .. (131.89,173.12) ;  
\draw  [draw opacity=0] (110.74,161.59) .. controls (110.74,161.6) and (110.75,161.6) .. (110.75,161.61) .. controls (116.08,165.85) and (122.9,166.83) .. (129.47,164.96) -- (127.02,138.66) -- cycle ; \draw   (110.74,161.59) .. controls (110.74,161.6) and (110.75,161.6) .. (110.75,161.61) .. controls (116.08,165.85) and (122.9,166.83) .. (129.47,164.96) ;  
\draw  [color={rgb, 255:red, 0; green, 0; blue, 0 }  ,draw opacity=1 ][fill={rgb, 255:red, 208; green, 2; blue, 27 }  ,fill opacity=1 ][line width=0.75]  (102.91,147.31) .. controls (101.74,147.33) and (100.76,144.86) .. (100.73,141.8) .. controls (100.7,138.74) and (101.63,136.24) .. (102.81,136.22) .. controls (103.98,136.2) and (104.95,138.66) .. (104.98,141.73) .. controls (105.01,144.79) and (104.08,147.29) .. (102.91,147.31) -- cycle ;
\draw  [color={rgb, 255:red, 0; green, 0; blue, 0 }  ,draw opacity=1 ][fill={rgb, 255:red, 208; green, 2; blue, 27 }  ,fill opacity=1 ][line width=0.75]  (102.8,129.5) .. controls (101.63,129.52) and (100.66,127.05) .. (100.63,123.99) .. controls (100.6,120.93) and (101.53,118.43) .. (102.7,118.41) .. controls (103.87,118.39) and (104.85,120.85) .. (104.88,123.92) .. controls (104.91,126.98) and (103.98,129.48) .. (102.8,129.5) -- cycle ;
\draw  [color={rgb, 255:red, 0; green, 0; blue, 0 }  ,draw opacity=1 ][fill={rgb, 255:red, 208; green, 2; blue, 27 }  ,fill opacity=1 ][line width=0.75]  (106.23,132.86) .. controls (106.21,131.7) and (108.69,130.73) .. (111.78,130.7) .. controls (114.87,130.67) and (117.39,131.59) .. (117.41,132.76) .. controls (117.43,133.92) and (114.95,134.88) .. (111.86,134.91) .. controls (108.77,134.94) and (106.25,134.02) .. (106.23,132.86) -- cycle ;
\draw  [color={rgb, 255:red, 0; green, 0; blue, 0 }  ,draw opacity=1 ][fill={rgb, 255:red, 208; green, 2; blue, 27 }  ,fill opacity=1 ][line width=0.75]  (88.2,132.96) .. controls (88.18,131.8) and (90.66,130.83) .. (93.75,130.81) .. controls (96.84,130.78) and (99.36,131.7) .. (99.38,132.86) .. controls (99.4,134.02) and (96.92,134.99) .. (93.83,135.02) .. controls (90.74,135.04) and (88.22,134.12) .. (88.2,132.96) -- cycle ;
\draw  [color={rgb, 255:red, 0; green, 0; blue, 0 }  ,draw opacity=1 ][fill={rgb, 255:red, 208; green, 2; blue, 27 }  ,fill opacity=1 ][line width=0.75]  (100.22,135.15) .. controls (101.07,135.95) and (100.02,138.38) .. (97.87,140.58) .. controls (95.72,142.78) and (93.29,143.91) .. (92.44,143.11) .. controls (91.59,142.3) and (92.64,139.87) .. (94.79,137.67) .. controls (96.94,135.47) and (99.37,134.34) .. (100.22,135.15) -- cycle ;
\draw  [color={rgb, 255:red, 0; green, 0; blue, 0 }  ,draw opacity=1 ][fill={rgb, 255:red, 208; green, 2; blue, 27 }  ,fill opacity=1 ][line width=0.75]  (100.19,130.26) .. controls (99.38,131.1) and (96.93,130.06) .. (94.71,127.93) .. controls (92.5,125.79) and (91.36,123.38) .. (92.17,122.54) .. controls (92.98,121.7) and (95.43,122.74) .. (97.65,124.88) .. controls (99.86,127.01) and (101,129.42) .. (100.19,130.26) -- cycle ;
\draw  [color={rgb, 255:red, 0; green, 0; blue, 0 }  ,draw opacity=1 ][fill={rgb, 255:red, 208; green, 2; blue, 27 }  ,fill opacity=1 ][line width=0.75]  (105.27,135.3) .. controls (104.42,136.1) and (105.48,138.53) .. (107.63,140.73) .. controls (109.78,142.92) and (112.21,144.06) .. (113.06,143.25) .. controls (113.91,142.45) and (112.85,140.02) .. (110.7,137.82) .. controls (108.55,135.62) and (106.12,134.49) .. (105.27,135.3) -- cycle ;
\draw  [color={rgb, 255:red, 0; green, 0; blue, 0 }  ,draw opacity=1 ][fill={rgb, 255:red, 208; green, 2; blue, 27 }  ,fill opacity=1 ][line width=0.75]  (105.27,130.08) .. controls (104.42,129.28) and (105.48,126.85) .. (107.63,124.65) .. controls (109.78,122.45) and (112.21,121.32) .. (113.06,122.13) .. controls (113.91,122.93) and (112.85,125.36) .. (110.7,127.56) .. controls (108.55,129.76) and (106.12,130.89) .. (105.27,130.08) -- cycle ;
\draw  [color={rgb, 255:red, 208; green, 2; blue, 27 }  ,draw opacity=1 ][fill={rgb, 255:red, 208; green, 2; blue, 27 }  ,fill opacity=1 ][line width=0.75]  (99.38,132.86) .. controls (99.38,131) and (100.92,129.5) .. (102.81,129.5) .. controls (104.7,129.5) and (106.23,131) .. (106.23,132.86) .. controls (106.23,134.72) and (104.7,136.22) .. (102.81,136.22) .. controls (100.92,136.22) and (99.38,134.72) .. (99.38,132.86) -- cycle ;

\draw (170,225) node [anchor=north west][inner sep=0.75pt]    {$\mathcal{H}^{A_{I}}$};
\draw (165.4,70) node [anchor=north west][inner sep=0.75pt]    {$\mathcal{H}^{A_{O}}$};
\draw (496.4,70) node [anchor=north west][inner sep=0.75pt]    {$\mathcal{H}^{B_{O}}$};
\draw (497.07,225) node [anchor=north west][inner sep=0.75pt]    {$\mathcal{H}^{B_{I}}$};
\draw (186,141.67) node [anchor=north west][inner sep=0.75pt]  [font=\Large,color={rgb, 255:red, 208; green, 2; blue, 27 }  ,opacity=1 ]  {$M_{a|x}$};
\draw (457.33,143) node [anchor=north west][inner sep=0.75pt]  [font=\Large,color={rgb, 255:red, 74; green, 144; blue, 226 }  ,opacity=1 ]  {$M_{b|y}$};
\draw (144,184.67) node [anchor=north west][inner sep=0.75pt]  [font=\Large]  {$x$};
\draw (326,136) node [anchor=north west, inner sep=0.75pt, font=\huge, color={rgb, 255:red, 126; green, 211; blue, 33 }, opacity=1, text=black]  {$W$};
\draw (144.67,111.33) node [anchor=north west][inner sep=0.75pt]  [font=\Large]  {$a$};
\draw (552.33,104.33) node [anchor=north west][inner sep=0.75pt]  [font=\Large]  {$b$};
\draw (554.67,185.33) node [anchor=north west][inner sep=0.75pt]  [font=\Large]  {$y$};

  \draw  [color={rgb, 255:red, 255; green, 255; blue, 255 }  ,draw opacity=1 ][fill={rgb, 255:red, 255; green, 255; blue, 255 }  ,fill opacity=1 ] (282.8,34) -- (437,37.94) -- (436.7,60.68) -- (282.8,60.7) -- cycle ;

  \draw  [color={rgb, 255:red, 255; green, 255; blue, 255 }  ,draw opacity=1 ][fill={rgb, 255:red, 255; green, 255; blue, 255 }  ,fill opacity=1 ] (282.7,260) -- (436.5,265.56) -- (436.8,280) -- (282.7,280.73) -- cycle ;
\end{tikzpicture}

%% file: images/measure_and_reprepare_channel.tex
\tikzset{every picture/.style={line width=0.75pt}} 

\begin{tikzpicture}[x=0.75pt,y=0.75pt,yscale=-1,xscale=1,scale=1]

\draw[ultra thick, color={rgb, 255:red, 208; green, 2; blue, 27 }]   (181.5,117) -- (269.5,117) -- (269.5,205) -- (181.5,205) -- cycle ;
\draw[ultra thick]    (138.67,179.67) -- (170.42,179.67) ;
\draw [shift={(173.42,179.67)}, rotate = 180] [fill={rgb, 255:red, 0; green, 0; blue, 0 }  ][line width=0.08]  [draw opacity=0] (5.36,-2.57) -- (0,0) -- (5.36,2.57) -- cycle    ;
\draw[ultra thick]    (139.5,141.5) -- (171.25,141.5) ;
\draw [shift={(136.5,141.5)}, rotate = 0] [fill={rgb, 255:red, 0; green, 0; blue, 0 }  ][line width=0.08]  [draw opacity=0] (5.36,-2.57) -- (0,0) -- (5.36,2.57) -- cycle    ;
\draw[ultra thick]    (226.8,110.8) -- (226.8,61) ;
\draw [shift={(226.8,58)}, rotate = 90] [fill={rgb, 255:red, 0; green, 0; blue, 0 }  ][line width=0.08]  [draw opacity=0] (5.36,-2.57) -- (0,0) -- (5.36,2.57) -- cycle    ;
\draw[ultra thick]    (226.8,264.8) -- (226.8,215) ;
\draw [shift={(226.8,212)}, rotate = 90] [fill={rgb, 255:red, 0; green, 0; blue, 0 }  ][line width=0.08]  [draw opacity=0] (5.36,-2.57) -- (0,0) -- (5.36,2.57) -- cycle    ;
\draw  [draw opacity=0] (110.74,161.6) .. controls (110.74,161.61) and (110.75,161.61) .. (110.75,161.62) .. controls (115.4,176.38) and (125.88,186.85) .. (137.17,189.58) -- (138.9,151.65) -- cycle ;

\draw (145,230) node [anchor=north west, inner sep=0.75pt, node font=\LARGE]    {$\mathcal{H}^{A_{I}}$};
\draw (140,70) node [anchor=north west, inner sep=0.75pt, node font=\LARGE]    {$\mathcal{H}^{A_{O}}$};
\draw (160,142.5) node [anchor=north west, inner sep=0.75pt,color={rgb, 255:red, 208; green, 2; blue, 27 }, node font=\LARGE] {$M_{a\vert x}^{A_I A_O}$};
\draw (496.4,66.6);
\draw (497.07,231.93);
\draw (186,141.67);
\draw (457.33,143);
\draw (115,175) node [anchor=north west, inner sep=0.75pt, font=\LARGE]  {$x$};
\draw (326,136);
\draw (115,135) node [anchor=north west, inner sep=0.75pt, font=\LARGE]  {$a$};
\draw (552.33,104.33);
\draw (554.67,185.33);

  \draw[ultra thick] (367.88,82.99) rectangle (504.67,241.62);
  \node[draw, shape=semicircle, minimum width=82pt, ultra thick, color={rgb, 255:red, 208; green, 2; blue, 27 }, fill=red!90!black!79!violet, fill opacity=0.65] (node4) at (436.27,211.77) {};
  \node[draw, shape=semicircle, minimum width=82pt, ultra thick, color={rgb, 255:red, 208; green, 2; blue, 27 }, rotate=180, fill=red!90!black!79!violet, fill opacity=0.65] (node5) at (436.27,112) {};
  \draw[ultra thick,color={rgb, 255:red, 208; green, 2; blue, 27 }] (node5.north) -- (node4.north);
  \draw[dashed] (360.4,82.99) -- (276.34,113.81);
  \draw[dashed] (360.4,241.62) -- (275.6,206.77);

\draw (390,97.5) node [anchor=north west, inner sep=0.75pt, color=black, node font=\LARGE]    {$\textcolor{white}{\rho_{a}^{A_O}}$};
\draw (385,190) node [anchor=north west, inner sep=0.75pt, color=black, node font=\LARGE]    {$\textcolor{white}{E_{a\vert x}^{A_I}}$};
\end{tikzpicture}

%% file: images/input_signalling.tex
\tikzset{every picture/.style={line width=0.75pt}} 

\begin{tikzpicture}[x=0.75pt,y=0.75pt,yscale=-1,xscale=1]

\draw  [color={rgb, 255:red, 65; green, 117; blue, 5 }  ,draw opacity=1 ][fill={rgb, 255:red, 65; green, 117; blue, 5 }  ,fill opacity=0.3 ] (56.8,137.24) .. controls (56.8,122.42) and (68.82,110.4) .. (83.64,110.4) .. controls (98.46,110.4) and (110.48,122.42) .. (110.48,137.24) .. controls (110.48,152.06) and (98.46,164.08) .. (83.64,164.08) .. controls (68.82,164.08) and (56.8,152.06) .. (56.8,137.24) -- cycle ;
\draw  [color={rgb, 255:red, 74; green, 144; blue, 226 }  ,draw opacity=1 ][fill={rgb, 255:red, 74; green, 144; blue, 226 }  ,fill opacity=0.3 ] (150,20.2) -- (200,20.2) -- (200,70.2) -- (150,70.2) -- cycle ;
\draw  [color={rgb, 255:red, 65; green, 117; blue, 5 }  ,draw opacity=1 ][fill={rgb, 255:red, 65; green, 117; blue, 5 }  ,fill opacity=0.3 ] (239.8,137.24) .. controls (239.8,122.42) and (251.82,110.4) .. (266.64,110.4) .. controls (281.46,110.4) and (293.48,122.42) .. (293.48,137.24) .. controls (293.48,152.06) and (281.46,164.08) .. (266.64,164.08) .. controls (251.82,164.08) and (239.8,152.06) .. (239.8,137.24) -- cycle ;
\draw  [color={rgb, 255:red, 245; green, 166; blue, 35 }  ,draw opacity=1 ][fill={rgb, 255:red, 245; green, 166; blue, 35 }  ,fill opacity=0.3 ] (56.8,237.64) .. controls (56.8,222.82) and (68.82,210.8) .. (83.64,210.8) .. controls (98.46,210.8) and (110.48,222.82) .. (110.48,237.64) .. controls (110.48,252.46) and (98.46,264.48) .. (83.64,264.48) .. controls (68.82,264.48) and (56.8,252.46) .. (56.8,237.64) -- cycle ;
\draw  [color={rgb, 255:red, 245; green, 166; blue, 35 }  ,draw opacity=1 ][fill={rgb, 255:red, 245; green, 166; blue, 35 }  ,fill opacity=0.3 ] (239.8,237.64) .. controls (239.8,222.82) and (251.82,210.8) .. (266.64,210.8) .. controls (281.46,210.8) and (293.48,222.82) .. (293.48,237.64) .. controls (293.48,252.46) and (281.46,264.48) .. (266.64,264.48) .. controls (251.82,264.48) and (239.8,252.46) .. (239.8,237.64) -- cycle ;
\draw    (83.76,208.12) -- (83.76,169.52) ;
\draw [shift={(83.76,166.52)}, rotate = 90] [fill={rgb, 255:red, 0; green, 0; blue, 0 }  ][line width=0.08]  [draw opacity=0] (7.14,-3.43) -- (0,0) -- (7.14,3.43) -- cycle    ;
\draw    (266.76,208.12) -- (266.76,169.52) ;
\draw [shift={(266.76,166.52)}, rotate = 90] [fill={rgb, 255:red, 0; green, 0; blue, 0 }  ][line width=0.08]  [draw opacity=0] (7.14,-3.43) -- (0,0) -- (7.14,3.43) -- cycle    ;
\draw    (146.16,74.6) -- (104.77,112.57) ;
\draw [shift={(102.56,114.6)}, rotate = 317.47] [fill={rgb, 255:red, 0; green, 0; blue, 0 }  ][line width=0.08]  [draw opacity=0] (7.14,-3.43) -- (0,0) -- (7.14,3.43) -- cycle    ;
\draw    (204.16,74.92) -- (244.77,112.87) ;
\draw [shift={(246.96,114.92)}, rotate = 223.06] [fill={rgb, 255:red, 0; green, 0; blue, 0 }  ][line width=0.08]  [draw opacity=0] (7.14,-3.43) -- (0,0) -- (7.14,3.43) -- cycle    ;
\draw [color={rgb, 255:red, 128; green, 128; blue, 128 }  ,draw opacity=1 ] [dash pattern={on 4.5pt off 4.5pt}]  (239.56,220.12) -- (115.2,153.46) ;
\draw [shift={(112.56,152.04)}, rotate = 28.19] [fill={rgb, 255:red, 128; green, 128; blue, 128 }  ,fill opacity=1 ][line width=0.08]  [draw opacity=0] (5.36,-2.57) -- (0,0) -- (5.36,2.57) -- cycle    ;
\draw [color={rgb, 255:red, 128; green, 128; blue, 128 }  ,draw opacity=1 ] [dash pattern={on 4.5pt off 4.5pt}]  (110.96,219.24) -- (236.3,153.43) ;
\draw [shift={(238.96,152.04)}, rotate = 152.3] [fill={rgb, 255:red, 128; green, 128; blue, 128 }  ,fill opacity=1 ][line width=0.08]  [draw opacity=0] (5.36,-2.57) -- (0,0) -- (5.36,2.57) -- cycle    ;

\draw (162.27,30.97) node [anchor=north west][inner sep=0.75pt]  [font=\Huge]  {$\Lambda $};
\draw (66.2,225.24) node [anchor=north west][inner sep=0.75pt]  [font=\Huge]  {$X$};
\draw (254.2,225.64) node [anchor=north west][inner sep=0.75pt]  [font=\Huge]  {$Y$};
\draw (67.6,121.36) node [anchor=north west][inner sep=0.75pt]  [font=\Huge]  {$A$};
\draw (252,123.56) node [anchor=north west][inner sep=0.75pt]  [font=\Huge]  {$B$};
\draw (98.58,206.27) node [anchor=north west][inner sep=0.75pt]  [font=\Large,color={rgb, 255:red, 128; green, 128; blue, 128 }  ,opacity=1 ,rotate=-331.4]  {$A\prec B$};
\draw (196.92,176.64) node [anchor=north west][inner sep=0.75pt]  [font=\Large,color={rgb, 255:red, 128; green, 128; blue, 128 }  ,opacity=1 ,rotate=-28.77]  {$B\prec A$};

\end{tikzpicture}

%% file: images/output_signalling.tex
\tikzset{every picture/.style={line width=0.75pt}} 

\begin{tikzpicture}[x=0.75pt,y=0.75pt,yscale=-1,xscale=1]

\draw  [color={rgb, 255:red, 65; green, 117; blue, 5 }  ,draw opacity=1 ][fill={rgb, 255:red, 65; green, 117; blue, 5 }  ,fill opacity=0.3 ] (56.8,137.24) .. controls (56.8,122.42) and (68.82,110.4) .. (83.64,110.4) .. controls (98.46,110.4) and (110.48,122.42) .. (110.48,137.24) .. controls (110.48,152.06) and (98.46,164.08) .. (83.64,164.08) .. controls (68.82,164.08) and (56.8,152.06) .. (56.8,137.24) -- cycle ;
\draw  [color={rgb, 255:red, 74; green, 144; blue, 226 }  ,draw opacity=1 ][fill={rgb, 255:red, 74; green, 144; blue, 226 }  ,fill opacity=0.3 ] (150,20.2) -- (200,20.2) -- (200,70.2) -- (150,70.2) -- cycle ;
\draw  [color={rgb, 255:red, 65; green, 117; blue, 5 }  ,draw opacity=1 ][fill={rgb, 255:red, 65; green, 117; blue, 5 }  ,fill opacity=0.3 ] (239.8,137.24) .. controls (239.8,122.42) and (251.82,110.4) .. (266.64,110.4) .. controls (281.46,110.4) and (293.48,122.42) .. (293.48,137.24) .. controls (293.48,152.06) and (281.46,164.08) .. (266.64,164.08) .. controls (251.82,164.08) and (239.8,152.06) .. (239.8,137.24) -- cycle ;
\draw  [color={rgb, 255:red, 245; green, 166; blue, 35 }  ,draw opacity=1 ][fill={rgb, 255:red, 245; green, 166; blue, 35 }  ,fill opacity=0.3 ] (56.8,237.64) .. controls (56.8,222.82) and (68.82,210.8) .. (83.64,210.8) .. controls (98.46,210.8) and (110.48,222.82) .. (110.48,237.64) .. controls (110.48,252.46) and (98.46,264.48) .. (83.64,264.48) .. controls (68.82,264.48) and (56.8,252.46) .. (56.8,237.64) -- cycle ;
\draw  [color={rgb, 255:red, 245; green, 166; blue, 35 }  ,draw opacity=1 ][fill={rgb, 255:red, 245; green, 166; blue, 35 }  ,fill opacity=0.3 ] (239.8,237.64) .. controls (239.8,222.82) and (251.82,210.8) .. (266.64,210.8) .. controls (281.46,210.8) and (293.48,222.82) .. (293.48,237.64) .. controls (293.48,252.46) and (281.46,264.48) .. (266.64,264.48) .. controls (251.82,264.48) and (239.8,252.46) .. (239.8,237.64) -- cycle ;
\draw    (83.76,208.12) -- (83.76,169.52) ;
\draw [shift={(83.76,166.52)}, rotate = 90] [fill={rgb, 255:red, 0; green, 0; blue, 0 }  ][line width=0.08]  [draw opacity=0] (7.14,-3.43) -- (0,0) -- (7.14,3.43) -- cycle    ;
\draw    (266.76,209.12) -- (266.76,170.52) ;
\draw [shift={(266.76,167.52)}, rotate = 90] [fill={rgb, 255:red, 0; green, 0; blue, 0 }  ][line width=0.08]  [draw opacity=0] (7.14,-3.43) -- (0,0) -- (7.14,3.43) -- cycle    ;
\draw    (146.16,74.6) -- (104.77,112.57) ;
\draw [shift={(102.56,114.6)}, rotate = 317.47] [fill={rgb, 255:red, 0; green, 0; blue, 0 }  ][line width=0.08]  [draw opacity=0] (7.14,-3.43) -- (0,0) -- (7.14,3.43) -- cycle    ;
\draw    (204.16,74.92) -- (244.77,112.87) ;
\draw [shift={(246.96,114.92)}, rotate = 223.06] [fill={rgb, 255:red, 0; green, 0; blue, 0 }  ][line width=0.08]  [draw opacity=0] (7.14,-3.43) -- (0,0) -- (7.14,3.43) -- cycle    ;
\draw [color={rgb, 255:red, 128; green, 128; blue, 128 }  ,draw opacity=1 ] [dash pattern={on 4.5pt off 4.5pt}]  (232.4,150.47) -- (119.4,149.49) ;
\draw [shift={(116.4,149.47)}, rotate = 0.49] [fill={rgb, 255:red, 128; green, 128; blue, 128 }  ,fill opacity=1 ][line width=0.08]  [draw opacity=0] (7.14,-3.43) -- (0,0) -- (7.14,3.43) -- cycle    ;
\draw [color={rgb, 255:red, 128; green, 128; blue, 128 }  ,draw opacity=1 ] [dash pattern={on 4.5pt off 4.5pt}]  (115.73,130.13) -- (230.73,130.13) ;
\draw [shift={(233.73,130.13)}, rotate = 180] [fill={rgb, 255:red, 128; green, 128; blue, 128 }  ,fill opacity=1 ][line width=0.08]  [draw opacity=0] (7.14,-3.43) -- (0,0) -- (7.14,3.43) -- cycle    ;

\draw (162.27,30.97) node [anchor=north west][inner sep=0.75pt]  [font=\Huge]  {$\Lambda $};
\draw (66.2,225.24) node [anchor=north west][inner sep=0.75pt]  [font=\Huge]  {$X$};
\draw (254.2,225.64) node [anchor=north west][inner sep=0.75pt]  [font=\Huge]  {$Y$};
\draw (67.6,121.36) node [anchor=north west][inner sep=0.75pt]  [font=\Huge]  {$A$};
\draw (252,123.56) node [anchor=north west][inner sep=0.75pt]  [font=\Huge]  {$B$};
\draw (143.7,105.27) node [anchor=north west][inner sep=0.75pt]  [font=\LARGE,color={rgb, 255:red, 128; green, 128; blue, 128 }  ,opacity=1 ,rotate=-359.84]  {$A\prec B$};
\draw (143.47,155) node [anchor=north west][inner sep=0.75pt]  [font=\LARGE,color={rgb, 255:red, 128; green, 128; blue, 128 }  ,opacity=1 ,rotate=-0.23]  {$B\prec A$};

\end{tikzpicture}

%% file: images/A_to_B_exc_graph.tex
\begin{figure}
\centering

\includegraphics[width=0.5\textwidth]{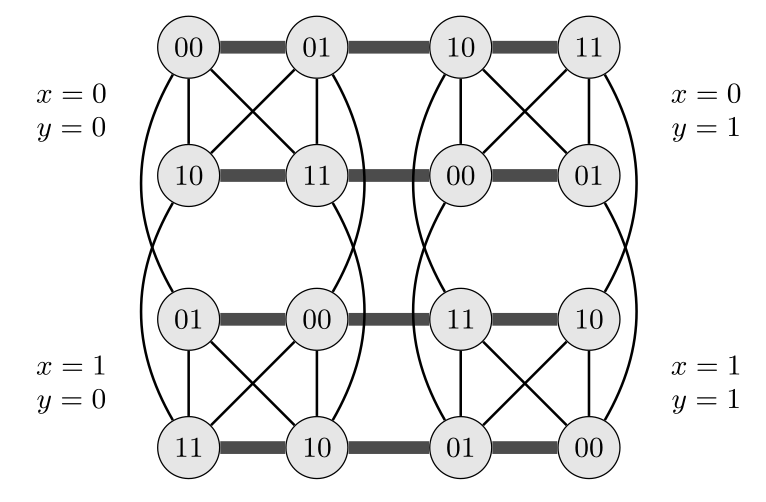} 
\caption{\textmd{Exclusivity graph for the $A \prec B$ order of the output-signaling scenario.} \label{fig:A_to_B_exc_graph}}

\end{figure}

%% file: images/boxes.tex
\tikzset{every picture/.style={line width=0.75pt}} 

\begin{tikzpicture}[x=0.75pt,y=0.75pt,yscale=-1,xscale=1,scale=0.65]

\draw  [color={rgb, 255:red, 245; green, 166; blue, 35 }  ,draw opacity=1 ][fill={rgb, 255:red, 245; green, 166; blue, 35 }  ,fill opacity=0.4 ][line width=1.5]  (180.08,100) .. controls (180.42,99.8) and (226.68,44.72) .. (329.28,43.72) .. controls (431.88,42.72) and (480.65,98.98) .. (482.32,99.72) .. controls (484,100.47) and (498.95,151.6) .. (500.16,154.55) .. controls (501.37,157.51) and (516.79,204.47) .. (518,208.83) .. controls (519.21,213.19) and (481.07,236.78) .. (472.88,261.32) .. controls (464.69,285.86) and (486.07,315.24) .. (482.32,319.32) .. controls (478.57,323.4) and (429.78,380.02) .. (329.28,379.72) .. controls (228.78,379.42) and (183.84,322.92) .. (180.08,319.6) .. controls (176.31,316.28) and (175.88,286.67) .. (169.28,262.12) .. controls (162.68,237.57) and (143.72,211.2) .. (144.4,209.1) .. controls (145.08,207.01) and (160.95,158.35) .. (162.2,154.35) .. controls (163.45,150.35) and (179.73,100.2) .. (180.08,100) -- cycle ;
\draw  [fill={rgb, 255:red, 255; green, 255; blue, 255 }  ,fill opacity=1 ] (482.32,100) -- (518,209.66) -- (482.32,319.6) -- (331.2,319.32) -- (180.08,319.6) -- (144.4,209.94) -- (180.08,100) -- cycle ;
\draw  [color={rgb, 255:red, 144; green, 19; blue, 254 }  ,draw opacity=1 ][fill={rgb, 255:red, 144; green, 19; blue, 254 }  ,fill opacity=0.3 ][line width=1.5]  (482.32,99.72) -- (518,209.38) -- (482.32,319.32) -- (331.2,319.04) -- (180.08,319.32) -- (144.4,209.66) -- (174.6,116.61) -- (180.08,99.72) -- cycle ;
\draw [color={rgb, 255:red, 144; green, 19; blue, 254 }  ,draw opacity=1 ][line width=1.5]  [dash pattern={on 5.63pt off 4.5pt}]  (180.08,319.04) -- (482.32,319.04) ;
\draw  [color={rgb, 255:red, 74; green, 144; blue, 226 }  ,draw opacity=1 ][dash pattern={on 5.63pt off 4.5pt}][line width=1.5]  (241.5,139.5) .. controls (335.5,99.5) and (478.71,100.5) .. (479.6,100) .. controls (480.5,99.5) and (432,121.5) .. (469.5,144.5) .. controls (507,167.5) and (515.5,200.13) .. (518,209.1) .. controls (520.5,218.08) and (481.07,237.06) .. (472.88,261.6) .. controls (464.69,286.14) and (486.07,315.52) .. (482.32,319.6) .. controls (478.57,323.68) and (407.5,362.05) .. (330,362.3) .. controls (252.5,362.55) and (183.84,323.2) .. (180.08,319.88) .. controls (176.31,316.55) and (175.88,286.95) .. (169.28,262.4) .. controls (162.68,237.85) and (142.75,209.25) .. (144.4,209.38) .. controls (146.05,209.52) and (147.5,179.5) .. (241.5,139.5) -- cycle ;
\draw  [fill={rgb, 255:red, 255; green, 255; blue, 255 }  ,fill opacity=1 ] (482.32,100) -- (518,208.83) -- (482.32,319.04) -- (180.08,319.04) -- (144.4,209.1) -- cycle ;
\draw  [color={rgb, 255:red, 208; green, 2; blue, 27 }  ,draw opacity=1 ][fill={rgb, 255:red, 208; green, 2; blue, 27 }  ,fill opacity=0.3 ][line width=1.5]  (482.32,100) -- (518,209.1) -- (482.32,319.32) -- (180.08,319.32) -- (144.4,209.38) -- cycle ;
\draw  [fill={rgb, 255:red, 144; green, 19; blue, 254 }  ,fill opacity=1 ] (141.68,209.66) .. controls (141.68,208.16) and (142.9,206.94) .. (144.4,206.94) .. controls (145.9,206.94) and (147.12,208.16) .. (147.12,209.66) .. controls (147.12,211.16) and (145.9,212.38) .. (144.4,212.38) .. controls (142.9,212.38) and (141.68,211.16) .. (141.68,209.66) -- cycle ;
\draw  [fill={rgb, 255:red, 144; green, 19; blue, 254 }  ,fill opacity=1 ] (515.28,208.83) .. controls (515.28,207.32) and (516.5,206.11) .. (518,206.11) .. controls (519.5,206.11) and (520.72,207.32) .. (520.72,208.83) .. controls (520.72,210.33) and (519.5,211.55) .. (518,211.55) .. controls (516.5,211.55) and (515.28,210.33) .. (515.28,208.83) -- cycle ;
\draw [color={rgb, 255:red, 144; green, 19; blue, 254 }  ,draw opacity=1 ][line width=1.5]  [dash pattern={on 5.63pt off 4.5pt}]  (180.08,319.32) -- (482.32,319.32) ;
\draw  [fill={rgb, 255:red, 144; green, 19; blue, 254 }  ,fill opacity=1 ] (479.6,319.6) .. controls (479.6,318.1) and (480.82,316.88) .. (482.32,316.88) .. controls (483.83,316.88) and (485.04,318.1) .. (485.04,319.6) .. controls (485.04,321.1) and (483.83,322.32) .. (482.32,322.32) .. controls (480.82,322.32) and (479.6,321.1) .. (479.6,319.6) -- cycle ;
\draw  [fill={rgb, 255:red, 144; green, 19; blue, 254 }  ,fill opacity=1 ] (177.36,319.6) .. controls (177.36,318.1) and (178.57,316.88) .. (180.08,316.88) .. controls (181.58,316.88) and (182.8,318.1) .. (182.8,319.6) .. controls (182.8,321.1) and (181.58,322.32) .. (180.08,322.32) .. controls (178.57,322.32) and (177.36,321.1) .. (177.36,319.6) -- cycle ;
\draw [color={rgb, 255:red, 144; green, 19; blue, 254 }  ,draw opacity=1 ]   (370.2,98.75) .. controls (366.68,88.41) and (360.64,87.94) .. (353.69,87.71) ;
\draw [shift={(350.8,87.6)}, rotate = 2.81] [fill={rgb, 255:red, 144; green, 19; blue, 254 }  ,fill opacity=1 ][line width=0.08]  [draw opacity=0] (5.36,-2.57) -- (0,0) -- (5.36,2.57) -- cycle    ;
\draw [color={rgb, 255:red, 144; green, 19; blue, 254 }  ,draw opacity=1 ]   (278.2,320.4) .. controls (282.83,328.96) and (288.86,331.52) .. (300.11,332.41) ;
\draw [shift={(302.95,332.6)}, rotate = 183.06] [fill={rgb, 255:red, 144; green, 19; blue, 254 }  ,fill opacity=1 ][line width=0.08]  [draw opacity=0] (5.36,-2.57) -- (0,0) -- (5.36,2.57) -- cycle    ;
\draw  [fill={rgb, 255:red, 144; green, 19; blue, 254 }  ,fill opacity=1 ] (177.36,100) .. controls (177.36,98.5) and (178.57,97.28) .. (180.08,97.28) .. controls (181.58,97.28) and (182.8,98.5) .. (182.8,100) .. controls (182.8,101.5) and (181.58,102.72) .. (180.08,102.72) .. controls (178.57,102.72) and (177.36,101.5) .. (177.36,100) -- cycle ;
\draw  [fill={rgb, 255:red, 144; green, 19; blue, 254 }  ,fill opacity=1 ] (479.6,100) .. controls (479.6,98.5) and (480.82,97.28) .. (482.32,97.28) .. controls (483.83,97.28) and (485.04,98.5) .. (485.04,100) .. controls (485.04,101.5) and (483.83,102.72) .. (482.32,102.72) .. controls (480.82,102.72) and (479.6,101.5) .. (479.6,100) -- cycle ;

\draw (303.23,322) node [anchor=north west][inner sep=0.75pt]  [font=\small] [align=left] {$\tilde{\mathcal{I}}_\text{LGYNI}$};
\draw (295,73.92) node [anchor=north west][inner sep=0.75pt]  [font=\small] [align=left] {$\tilde{\mathcal{I}}_\text{GYNI}$};
\draw (180.08,105) node [anchor=north west][inner sep=0.75pt]  [font=\Huge,color={rgb, 255:red, 144; green, 19; blue, 254 }  ,opacity=1 ]  {$\mathcal{C}$};
\draw (350,200) node [anchor=north west][inner sep=0.75pt]  [font=\Huge,color={rgb, 255:red, 208; green, 2; blue, 27 }  ,opacity=1 ]  {$\mathcal{OS}$};

\end{tikzpicture}